\documentclass[traditabstract]{aa}
\usepackage{graphicx}
\usepackage{txfonts}
\usepackage{appendix}
\usepackage{color}
\begin{document}
   \title{\textit{Herschel}\thanks{Herschel is an ESA space observatory 
            with science instruments provided by European-led Principal 
            Investigator consortia and with important participation from NASA.}
            imaging and spectroscopy of the nebula around the luminous blue 
            variable star WRAY 15-751\thanks{Based in part on observations 
            collected at the European Southern Observatory, La Silla, Chile}
         }

   \author{C. Vamvatira-Nakou\inst{1}
          \and D. Hutsem\'{e}kers\inst{1}\fnmsep\thanks{Senior Research 
               Associate FNRS}
          \and P. Royer\inst{2} 
          \and Y. Naz\'e\inst{1}\fnmsep\thanks{Research Associate FNRS}
          \and P. Magain\inst{1}
          \and \\K. Exter\inst{2} 
          \and C. Waelkens\inst{2} 
          \and M. A. T. Groenewegen\inst{3}
          }

   \institute{Institut d'Astrophysique et de G\'{e}ophysique, Universit\'{e} de Li\`ege, All\'{e}e du 
              6 ao\^ut, 17 - B\^at. B5c, B-4000 Li\`ege, Belgium \\
              \email{vamvatira@astro.ulg.ac.be}
           \and Instituut voor Sterrenkunde, Katholieke Universiteit Leuven, Celestijnenlaan 200D, B-3001 Leuven, Belgium 
           \and Koninklijke Sterrenwacht van Belgi\"e, Ringlaan 3, B-1180 Brussels, Belgium
             }

   \date{Received , 2013; accepted , 2013}

 
\abstract{
We have obtained far-infrared {\it Herschel} PACS imaging and
spectroscopic observations of the nebular environment of the luminous
blue variable (LBV) WRAY 15-751.

The far-infrared images clearly show that the main, dusty nebula is a
shell of radius 0.5 pc and width 0.35 pc extending outside the
H$\alpha$ nebula.  Furthermore, these images reveal a second, bigger
and fainter dust nebula that is observed for the first time. Both nebulae
lie in an empty cavity, very likely the remnant of the O-star wind bubble
formed when the star was on the main sequence. The kinematic ages of
the nebulae are calculated to be about 2 $\times$ 10$^{4}$ and 8
$\times$ 10$^{4}$ years, and we estimated that each nebula contains
$\sim$ 0.05 M$_{\odot}$ of dust. Modeling of the inner nebula indicates
a Fe-rich dust.

The far-infrared spectrum of the main nebula revealed forbidden
emission lines coming from ionized and neutral gas. Our study
shows that the main nebula consists of a shell of ionized
gas surrounded by a thin photodissociation region illuminated by an
"average" early-B star. We derive the abundance ratios N/O = 1.0
$\pm$ 0.4 and C/O = 0.4 $\pm$ 0.2,  which indicate a mild N/O enrichment.
From both the ionized and neutral gas components we estimate that the
inner shell contains 1.7 $\pm$ 0.6 M$_{\odot}$ of gas. Assuming a
similar dust-to-gas ratio for the outer nebula, the total mass ejected
by WRAY 15-751 amounts to 4 $\pm$ 2 M$_{\odot}$.

The measured abundances, masses and kinematic ages of the nebulae were
used to constrain the evolution of the star and the epoch at which the
nebulae were ejected.  Our results point to an ejection of the nebulae
during the red super-giant (RSG) evolutionary phase of an $\sim$ 40 M$_{\odot}$ star.
The multiple shells around the star suggest that the
mass-loss was not a continuous ejection but rather a series of episodes of
extreme mass-loss.  Our measurements are compatible with the recent
evolutionary tracks computed for an $\sim$ 40 M$_{\odot}$ star with
little rotation.  They support the O--BSG--RSG--YSG--LBV filiation and
the idea that high-luminosity and low-luminosity LBVs follow different
evolutionary paths.

}

   \authorrunning{C. Vamvatira-Nakou et al.}
   \titlerunning{The nebula around the LBV star WRAY 15-751}

   \maketitle

%

\section{Introduction}
\label{sec:introduction}

Luminous blue variables (LBVs), or S Doradus variables, represent a
short stage ($\sim10^{4} - 10^{5}\ \mathrm{yr}$) in the evolution of
massive stars with initial mass $\ge 30\ \mathrm{M}_{\odot}$ (Maeder
\& Meynet \cite{maed10}). They are located in the upper left part of
the Hertzsprung-Russell diagram (HRD), although some of them undergo
occasional excursions to the right of the HRD. Their main
characteristics are a) photometric variability, from giant eruptions,
$\geq 2\ \mathrm{mag}$, to small oscillations, $\sim$0.1 mag; b) high
luminosity, $\sim10^{6}\ \mathrm{L}_{\odot}$; and c) high mass-loss
rates, $\sim10^{-5} - 10^{-4}\ \mathrm{M}_{\odot}\ \mathrm{yr}^{-1}$
(Humphreys \& Davidson \cite{hum94}).

According to current evolutionary scenarios (Maeder \& Meynet
\cite{maed10}), an early-type O star evolves into a wolf-rayet (WR) star by losing
a significant fraction of its initial mass. Progressively, the outer
layers of the star are removed, revealing a ``bare core'' that becomes a WR
star. One way to lose mass is through stellar winds. However, in the
past few years the mass-loss rates of O stars have been revised downward by
up to one order of magnitude (Fullerton et al. \cite{fullerton}) and more
often by a factor of a few (Bouret et al. \cite{bouret}; Puls et al.
\cite{puls}), highlighting the key role played by episodes of extreme
mass-loss in an intermediate evolutionary phase (LBV or red supergiant
phase).

Most LBVs are surrounded by ejected nebulae (Hutsem\'{e}kers
\cite{hut94}; Nota et al. \cite{nota1}). The \ion{H}{ii} nebulae have
diameters of 0.5-2 pc, expansion velocities of a few tens of
$\mathrm{km}\ \mathrm{s}^{-1}$, and dynamical ages of $3\times10^3$ to
$5\times10^4$ yr. Their morphologies are usually axisymmetric, from
mildly to extremely bipolar or elliptical. Previous infrared and
millimeter studies of LBV nebulae have revealed not only dust but also
molecular gas (CO) (McGregor et al. \cite{mcgregor}; Hutsem\'{e}kers
\cite{hut97}; Nota et al. \cite{nota2}).

There are many questions about the detailed evolution of these massive
stars. For instance, we still do not know when and how the nebulae are
ejected, what causes the strong mass-loss phases and what leads to the
giant eruptions observed in some of them. Also, important quantities
such as the nebular mass and the gas composition (CNO abundances) are
very uncertain.

WRAY 15-751 (= Hen 3-591 = IRAS 11065-6026) was first considered to be
a possible WR star by Henize (Roberts \cite{roberts}) because of a
perceptibly widened H$\alpha$ emission line. Carlson and
Henize (\cite{carlson}) included it in their sample of southern
peculiar emission-line stars and classified it as a Bep star on the
basis of the strong $[\ion{Fe}{ii}]$ emission lines characterizing its
spectrum. Based on a photometric and spectroscopic study in the
optical, Hu et al. (\cite{hu}) concluded that WRAY 15-751 is a
variable star with spectral type O9.5. After estimating its
distance ($r>5\ \mbox{kpc}$) and temperature ($T_\mathrm{eff}=30000\
\mathrm{K}$), these authors calculated a lower limit of log $L/L_{\odot}$ equal
to 5.7. By plotting these data in an evolutionary diagram, they
revealed that WRAY 15-751 is located in the region of LBV stars, with
a lower limit on the initial mass of approximately 50 $M_{\odot}$. De
Winter et al. (\cite{dewinter}) made an extensive comparative study of
the optical and ultraviolet characteristic of this star with those of
the LBVs AG Car and HR Car. They concluded that WRAY 15-751 was a LBV
in a phase of quiescence surrounded by a cold dusty circumstellar shell
with strong emission in the far-infrared, like HR Car.

Based on the available photometry, Sterken et al. (\cite{ste08})
showed that WRAY 15-751 exhibits strong variability, confirming that the star
belongs to the S Dor class. Its variations have an amplitude of about two
magnitudes in V and a cycle length of several decades, similar to the
observed variations of AG Car. The star moved from V $\simeq$ 12.5 and
$T_\mathrm{eff} \simeq 30000$ K in 1989, to V $\simeq$ 10.5 and $T_\mathrm{eff}
\simeq 9000$ K in 2008.

Hutsem\'{e}kers and Van Drom (\cite{hut91}, hereafter HVD) studied
WRAY 15-751 with optical photometric and spectroscopic data. They found
that the star is surrounded by a ring nebula of ionized gas
with a diameter of about 22$\arcsec$. The nebula appeared non-uniform
in brightness and is  apparently not detached from the central star. This
fact led them to conclude that the nebula might arise from a
continuous mass-loss instead of from a sudden outburst. Based on their
spectral analysis, they also suggested that the nebula is expanding
almost symmetrically at 26 km s$^{-1}$.

The first infrared study of the nebula around WRAY 15-751 was made by
Voors et al (\cite{voo00}). By modeling ground-based infrared images
taken at about 10\ \mbox{$\mu$m} and ISO spectroscopic observations,
they derived some properties of the circumstellar dust around the
star: the distribution of emission is roughly spherical, the dust shell
is detached and slightly elongated; there is neutral gas outside the
dust shell and ionized gas only in the inner part of it; the dust shell
contains on the average large grains and a minor population of warm very
small grains.

Weis (\cite{weis}) made a detailed kinematic and morphological study
of the nebula and found that, in addition to a nearly spherical shell, it
also displays a bipolar-like structure (caps). Duncan and White
(\cite{duncan}) studied this nebula at radio wavelengths (3 and 6 cm)
and confirmed the almost attached nebula surrounding
the central star. Moreover, the subtraction of the central star as a
point source revealed a two-component inner structure, which was
interpreted by the authors as a disk or torus, suggesting a possible
mass transfer from a companion star.

\begin{figure*}[!]
\resizebox{\hsize}{!}{\includegraphics*{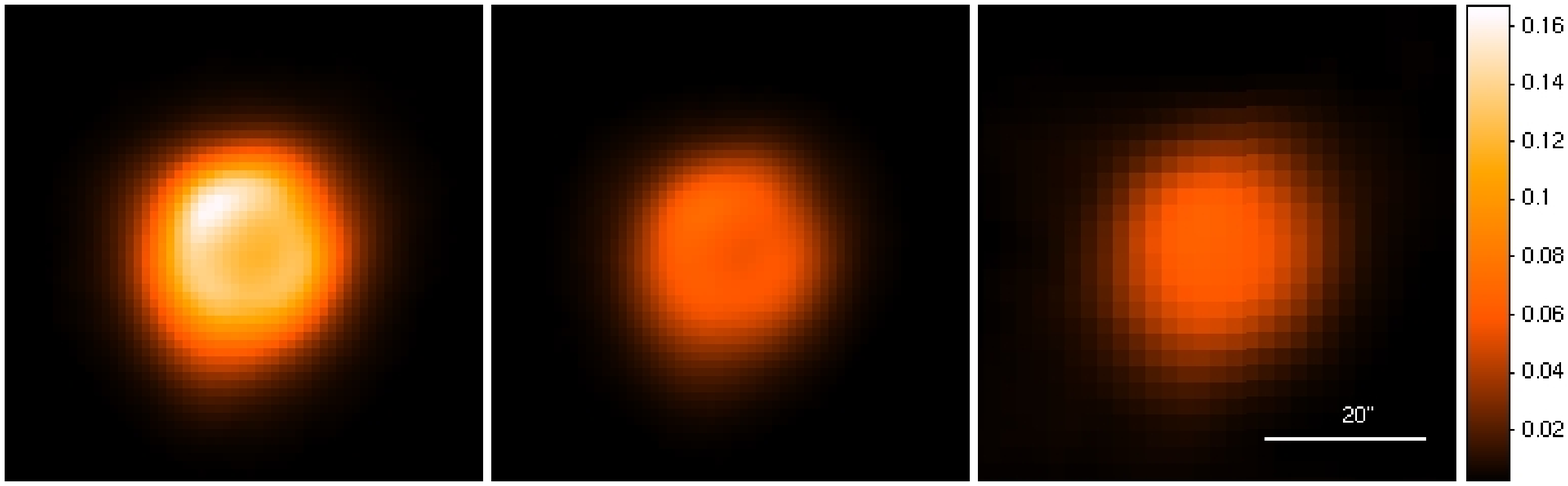}}\\%
\resizebox{\hsize}{!}{\includegraphics*{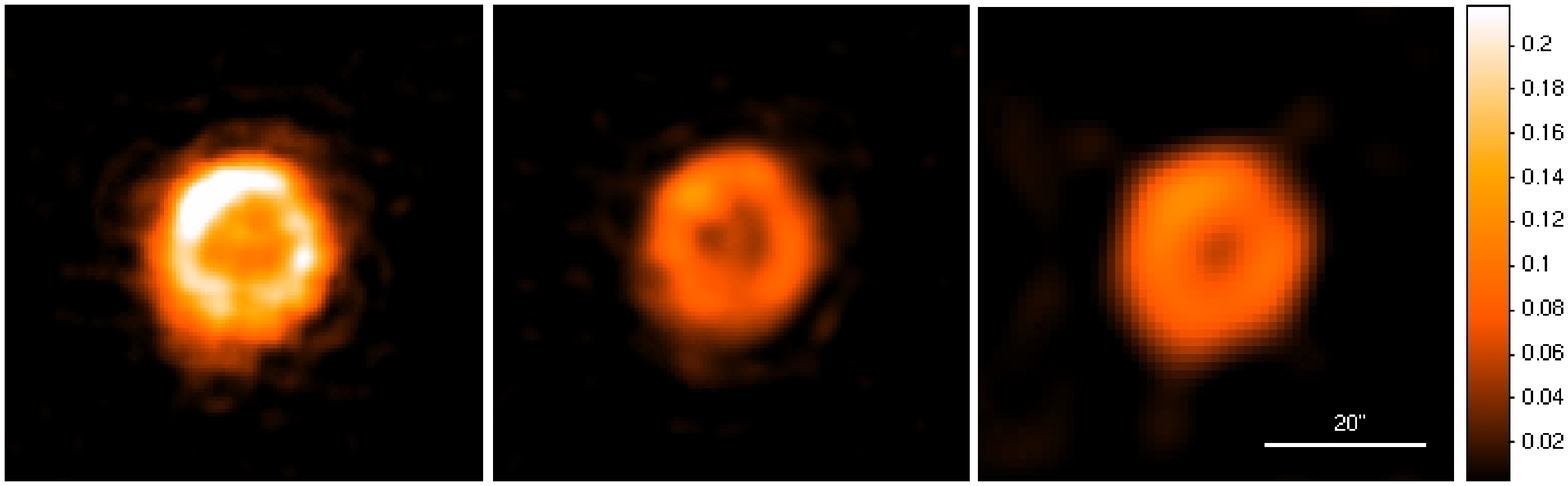}}
\caption{PACS images of the nebula around WRAY 15-751 at 70\
\mbox{$\mu$m}, 100\ \mbox{$\mu$m} and 160\ \mbox{$\mu$m}, from left to
right. Top: original images. Bottom: images deconvolved using the MCS
method. The size of each image is  1$\arcmin\times1\arcmin$. The scales
on the right correspond to the surface brightness (arbitrary units). North is
up and east is to the left.}
\label{imag}
\end{figure*}

Van Genderen et al. (\cite{vangenderen}) calculated a lower limit of 4-5 kpc
for the distance to WRAY 15-751, based on
the photometry of field stars. This value agrees with the suggestion of Hu et
al. (\cite{hu}). HVD found a larger distance of 7 kpc,
determined from the kinematics of the $[\ion{N}{ii}]$ lines. Pasquali
et al. (\cite{pasquali}) adopted a distance of 6 $\pm$ 1 kpc in their study
of the birth-cluster of WRAY 15-751 based on the radial velocity of the star
and its surrounding \ion{H}{ii} region. In this paper we adopt a distance
of 6 $\pm$ 1 kpc, which encompasses all measurements.

We analyze the images and the spectrum of the WRAY 15-751 nebula taken
by PACS (Photodetector Array Camera and Spectrometer, Poglitsch et al.
\cite{poglitsch}), one of the three instruments onboard the Herschel
Space Observatory (Pilbratt et al. \cite{pilbratt}). The paper is organized
as follows. The observations and the data reduction are presented in
Sect. 2. In Sect. 3 a description of the nebula far-infrared
morphology is given, while in Sect. 4 we give an overview of the spectrum. 
In Sect. 5 we model the dust continuum emission. The emission line spectrum
is presented and analyzed in Sect. 6. A general discussion follows in
Sect. 7 and conclusions appear in Sect. 8.

\section{Observations and data reduction}
\label{sec:observations and data reduction}

\subsection{Infrared observations}

The infrared imaging and spectroscopic observations were carried out
using PACS as part of the \textit{Mass-loss of Evolved StarS (MESS)}
Guaranteed Time Key Program (Groenewegen et al. \cite{groenewegen}).

The imaging observations of the WRAY 15-751 nebula were carried out on
January 2, 2010, which corresponds to the 233 observational day (OD) of
Herschel. The scan map mode was used. In this observing mode,
the telescope slews at constant speed ($20\arcsec/\mbox{s}$ in our
case) along parallel lines to cover the required area of the sky. For
each filter, two orthogonal scan maps were obtained so that our final
data set consists of maps at 70, 100 and 160\ \mbox{$\mu$m}. The
observation identification numbers (obsID) of the four scans are
1342188849, 1342188850, 1342188851, and 1342188852. The duration of
each one is 157s.

The data reduction was performed using the Herschel Interactive
Processing Environment (HIPE, Ott \cite{ott}). The `highpassFilter'
task was used to produce the final images as detailed in Groenewegen
et al.  (\cite{groenewegen}). The images were oversampled by a factor
of 3.2 with respect to the original pixel size, hence leading to
pixel sizes in the final maps of 1$\arcsec$ in the blue
(70, 100 $\mu$m) channel and 2$\arcsec$ in the red (160
$\mu$m) channel.  Since the highpassFilter task filters out the
largest structures, an independent data reduction was performed in all
three wavelengths using the
Microwave Anisotropy Dataset mapper (MADmap) algorithm (Cantalupo et
al.  \cite{cantalupo}) to investigate emission at large scales.  This
algorithm, also provided within HIPE, accounts for the significant
detector drift.

Deconvolution was applied to the three PACS images, produced with the
highpassFilter task, in an effort to
better reveal the morphology of the inner nebula.  For this purpose,
the point-spread functions (PSFs) of Vesta and the MCS deconvolution
method (Magain et al.  \cite{magain}) were used. The advantage of this
method is that it does not violate the sampling theorem.  Indeed, the
image is not deconvolved by the total PSF, which leads to an infinite
resolution, but the deconvolution makes use of a partial PSF chosen to
respect the desired resolution of the final deconvolved image. The
Herschel PACS PSF full widths at half maximum (FWHMs) are 5.2$\arcsec$,
7.7$\arcsec$ and 12$\arcsec$ at 70\ \mbox{$\mu$m}, 100\ \mbox{$\mu$m} and 160\
\mbox{$\mu$m}, respectively. After the deconvolution with the
corresponding PSF, the final spatial resolution is twice as good as the initial one.

The spectrum of the WRAY 15-751 nebula was taken on November 26, 2009
(OD 196) during the calibration phase of the instrument.  The PACS
integral-field spectrometer covers the wavelength range from 52\
\mbox{$\mu$m} to 220\ \mbox{$\mu$m} in two channels that operate
simultaneously in the blue, 52-98\ \mbox{$\mu$m} band (second order: B2A
52-73\ \mbox{$\mu$m} and B2B 70-105\ \mbox{$\mu$m}, 3rd order: B3A
52-73\ \mbox{$\mu$m}), and the red, 102-220\ \mbox{$\mu$m} band (first
order: R1A 133-220\ \mbox{$\mu$m} and R1B 102-203\ \mbox{$\mu$m}). It
has a resolving power of $\lambda/\delta\lambda \sim 940-5500$,
depending on the wavelength. It provides simultaneous imaging of a
$47\arcsec \times 47\arcsec$ field of view, resolved in $5\times5$
square spatial pixels (i.e., spaxels). An image slicer employing
reflective optics is used to re-arrange the two-dimensional
field-of-view along a $1\times25$ pixels entrance slit for the
gratings. We used the spectral energy distribution (SED) observing
template, which provides a complete coverage between 52 and 220\
\mbox{$\mu$m}. The two obsIDs of these observations are 1342187236 and
1342187237. The data reduction was also performed using HIPE, following
the standard data reduction steps, in particular the subtraction of
the background spectrum obtained through nodding.

\subsection{Visible observations}

The optical images of WRAY 15-751 and its nebula were obtained on
March 14, 1994, with the 3.6-m telescope at the European Southern
Observatory (ESO), La Silla, Chile. The EFOSC1 camera was used in its
coronographic mode: the 6$\arcsec$ circular coronographic mask was
inserted in the aperture wheel and positioned on the central star,
while the Lyot stop was inserted in the grism wheel (Melnick et
al. \cite{mel89}). A series of short (1s) and long (300s) exposures
were secured in a H$\alpha$+$[\ion{N}{ii}]$ filter ($\lambda_{\rm c}$
=6560.5\AA ; {\sc FWHM} =62.2\AA ), and in a continuum filter just
redward ($\lambda_{\rm c}$ = 6644.7\AA; {\sc FWHM} =61.0\AA ).  The
CCD pixel size was 0$\farcs$605 on the sky. The night was photometric
and the seeing around 1$\farcs$6.  The frames were bias-corrected and
flat-fielded. The continuum images were subtracted from the
H$\alpha$+$[\ion{N}{ii}]$ ones after correcting for the position
offsets and for the different filter transmissions, using field
stars. The resulting averaged images show more detail than those
displayed in HVD.  They can be compared to those obtained at the ESO
New Technology Telescope (NTT) with the STSci coronograph (Nota
\cite{not99}, Weis \cite{weis}).

\section{Morphology of the nebula}
\label{sec:morphology of the nebula}

\begin{figure}[t]
     \centering
     \includegraphics[width=5.9cm]{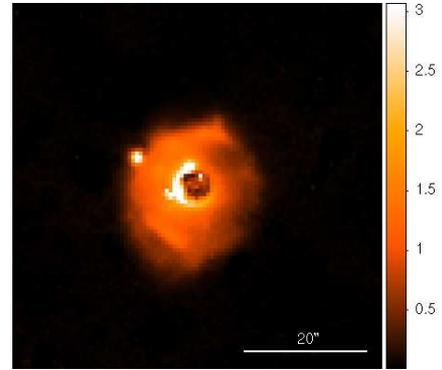}
     \caption{H$\alpha$+$[\ion{N}{ii}]$ image of the nebula around
      WRAY 15-751. The size of the image is 1$\arcmin \times 1
      \arcmin$. The scale on the right corresponds to  the surface brightness
      (arbitrary units). North is up and east is to the left. The
       central star is occulted by the coronograph spot.}
      \label{Halpha}
\end{figure}

The images of the WRAY 15-751 nebula at the three PACS wavelengths,
70\ \mbox{$\mu$m}, 100\ \mbox{$\mu$m} and 160\ \mbox{$\mu$m}, are
illustrated in Fig.~\ref{imag}. Note that the star is not
visible at these wavelengths. While the ionized gas does not appear
to be detached from the star (HVD; Duncan and White \cite{duncan};
Fig.~\ref{Halpha}), the dust emission seen in these images shows an
almost symmetric ring-like morphology, as suggested by Voors et
al. (\cite{voo00}) on the basis of mid-infrared imaging. This ring
shape is more clearly seen at 70\ \mbox{$\mu$m}, the wavelength at
which the spatial resolution is the highest.  The central part of the
nebula is clearly fainter than the ring. The very inner nebula, which is
unresolved in the optical but was detected at radio wavelengths by
Duncan and White (\cite{duncan}), is not seen in the PACS images.

\begin{figure}
\resizebox{\hsize}{!}{\includegraphics*{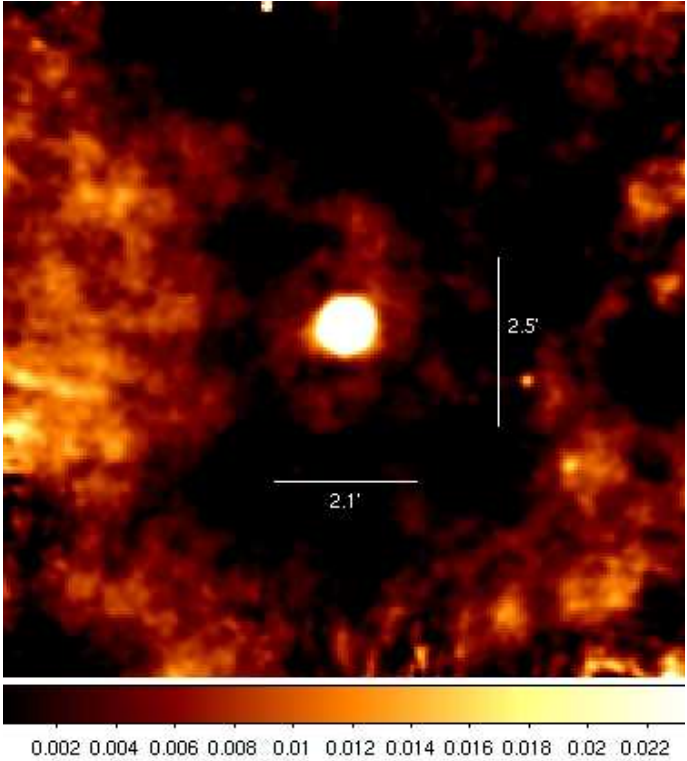}}
   \caption{PACS 100\ \mbox{$\mu$m} image of the nebula obtained
using the MADmap reduction algorithm. The size of the image is
10\arcmin$\times$10\arcmin. The scale on the bottom corresponds to  the
surface brightness (arbitrary units). North is up and east is to the left. A
faint extended elliptical nebulosity is seen around the bright
shell, the size of which is marked with the horizontal and vertical bars.
Both appear located inside a cavity in the interstellar
medium.}
    \label{2shell}
\end{figure}

\begin{figure}
\resizebox{\hsize}{!}{\includegraphics*{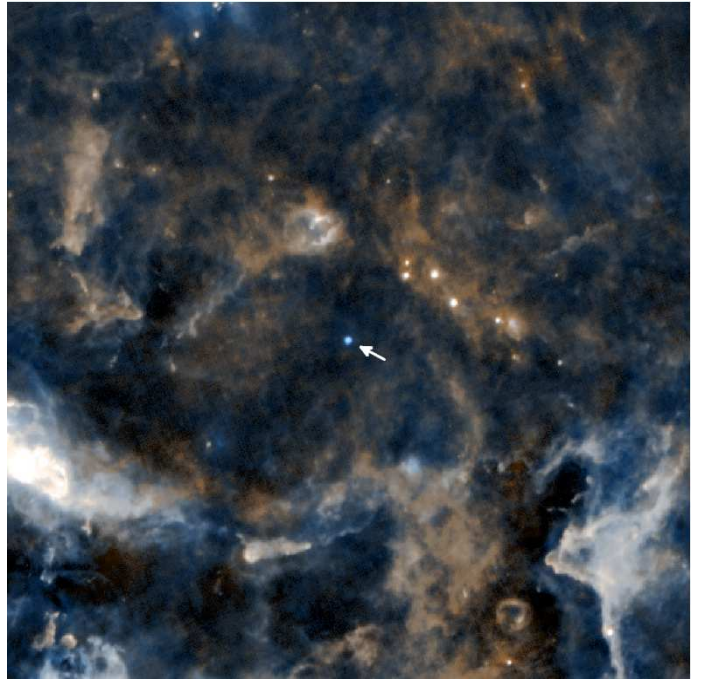}}
   \caption{Two-color (70 $\mu$m in blue and 160 $\mu$m in red)
image from the Hi-GAL survey of the complex environment of
WRAY 15-751 (the blue spot at the center of the image marked with a white arrow).  The size of
the image is 1$\degr$$\times$1$\degr$. North is up and east is to the
left.  A bubble $\sim$25$\arcmin$ in diameter and offset from the star
is tentatively seen around WRAY 15-751.}
    \label{higal}
\end{figure}

In the 70 $\mu$m deconvolved image, the ring extends up to $\sim18
\arcsec$ in radius with a width of $\sim12 \arcsec$.  Adopting a
distance of 6 kpc, these values corresponds to a nebular
radius of about 0.5 pc and to a ring width of 0.35 pc. The surface
brightness seems to be non-uniform, as the northeastern part of the
nebula is brighter than the other parts. This asymmetry in the
brightness distribution was also detected at optical and mid-infrared
wavelengths (HVD; Voors et al. \cite{voo00}).

Fig.~\ref{Halpha} illustrates the nebula around WRAY 15-751 in the
H$\alpha$+$[\ion{N}{ii}]$ light. The nebula essentially appears
disk-like with a circular rim 22$\arcsec$ in diameter, in agreement
with the measurement of HVD. It appears slightly elongated, with small
caps (Weis \cite{weis}) along the main axis (PA $\sim$ 155\degr, east
of north). The eastern part of the nebula is definitely brighter than
the western one. A similar morphology is observed at radio wavelengths
(Duncan and White \cite{duncan}). No diffuse emission can be detected
in the images obtained within the adjacent continuum filter.  The
H$\alpha$+$[\ion{N}{ii}]$ rim, which corresponds to the ionized gas
region, is inside the dust ring, which extends farther out.

In Fig.~\ref{2shell}, we illustrate the large-scale infrared emission
around WRAY 15-751, obtained after the reduction with the MADmap
algorithm. A much larger, very faint ellipsoidal nebula can be seen
circumscribing the WRAY 15-751 bright ring nebula. This outer nebula
is detected at all three wavelengths but it is more clearly seen at 100\
\mbox{$\mu$m}.  Its size is roughly 2.1$\arcmin \times 2.5\arcmin$,
which corresponds to a mean radius of 2 pc at a distance of 6 kpc. This
nebula is elongated along the same PA as the H$\alpha$+$[\ion{N}{ii}]$
inner shell (Fig.~\ref{Halpha}), supporting its physical association
to WRAY 15-751.  It is also interesting to note that it lies in a cavity,
probably cleaned up prior to the ejection of the nebula. The radius of
this empty cavity is about 4$\arcmin$, which corresponds to 7 pc at
a distance of 6 kpc.

The kinematic age of the two nebulae can be estimated, assuming that the
expansion velocity is the same in both cases. HVD measured the
expansion velocity to be ${\rm v}_{\mathrm{exp}} \sim$ 26 km
s$^{-1}$. Adopting this value, the inner nebula, of radius $r$ = 0.5
pc, has a kinematic age $t_{\mathrm{kin}} = r /{\rm v}_{\mathrm{exp}}$
of 1.9 $\times$ 10$^4$ years, while the outer nebula, of mean radius 2
pc, has a kinematic age of 7.5 $\times$ 10$^4$ years.

To explore the environment of WRAY 15-751 in more detail, we
considered the PACS observations of the field obtained in the
framework of the {\it Herschel} Infrared Galactic Plane survey
(Hi-GAL, Molinari et al. \cite{mol10}). The observations, made
immediately public for legacy, were retrieved from the archive
processed up to level 2.  The two orthogonal scans were added.

A two-color image is displayed in Fig.~\ref{higal}, illustrating the
complex interstellar environment around WRAY 15-751. In particular, we
can see a series of filaments that form a roughly circular structure
around WRAY 15-751. We tentatively interpret this structure as the
bubble formed by the O-star progenitor, although we cannot exclude 
a foreground/background structure. Velocity mapping would be needed
to ascertain the physical association. WRAY 15-751 appears to be offset with
respect to the bubble, possibly because of higher density
material northwest of the star. The radius of this bubble is about
12$\arcmin$, which corresponds to 20 pc at 6 kpc.

\section{Spectrum of the nebula: overview}
\label{sec:the spectrum of the nebula: overview}

\begin{figure}[t]
   \centering
   \includegraphics[width=7cm]{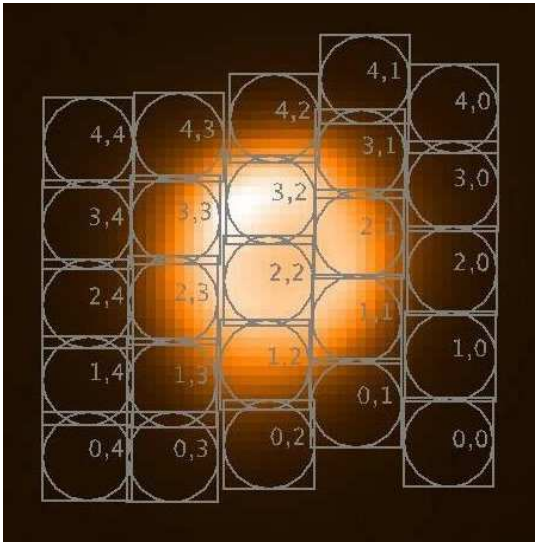}
   \caption{Footprint of the PACS spectral field of view on the image 
   of the nebula at 70\ \mbox{$\mu$m}. Each number pair is the label of a specific spaxel.
   The size of the image is $1\arcmin \times 1\arcmin$. North is up and east is to the left.}
   \label{wra_foot}
\end{figure}

The footprint of the PACS spectral field-of-view on the image of the
nebula at 70\ \mbox{$\mu$m} is shown in Fig.~\ref{wra_foot}. This
figure allows us to identify which spaxel corresponds to which part of
the nebula.  It must be noted that the whole inner ring nebula
is inside the spectral field of view although the center of the nebula is not
exactly at the central spaxel (2,2).

\begin{figure*}
\resizebox{\hsize}{!}{\includegraphics*{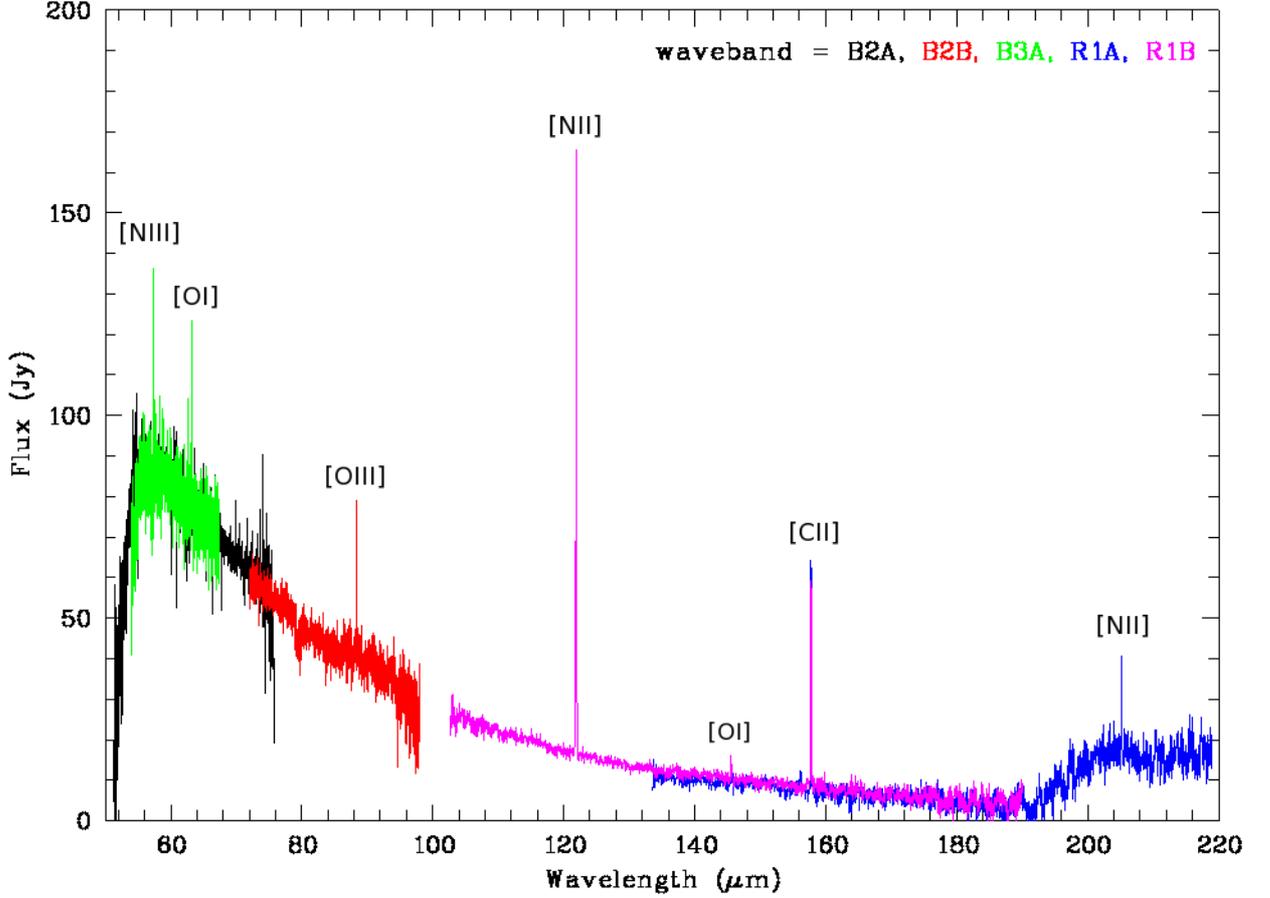}}
   \caption{PACS spectrum of WRAY 15-751, integrated over the nine
   central spaxels. Indicated are the lines [N{\sc iii}], [O{\sc i}], [O{\sc iii}],
   [N{\sc ii}] and [C{\sc ii}]. The continuum shape below 55 $\mu$m
   results from a yet-imperfect spectral response function correction,
   while above 190 $\mu$m it results from a light leak, from the second
   diffraction order of the grating in the first one. The different bands
   are indicated with different colors.}
   \label{wra_spec}
\end{figure*}

The spectrum of the nebula, integrated over the nine central spaxels,
is shown in Fig.~\ref{wra_spec}. The shape of the continuum below 55\
\mbox{$\mu$m} results from a yet-imperfect spectral response correction in
this range. Above 190\ \mbox{$\mu$m} the continuum shape results from a
light leak from the second diffraction order of the grating to the first one.

The following forbidden emission spectral lines are detected:
$[\ion{N}{iii}]$ $\lambda$ 57\ \mbox{$\mu$m}, $[\ion{O}{i}]$
$\lambda\lambda$ 63, 146\ \mbox{$\mu$m}, $[\ion{O}{iii}]$ $\lambda$
88\ \mbox{$\mu$m}, $[\ion{N}{ii}]$ $\lambda\lambda$ 122, 205\ \mbox{$\mu$m},
and $[\ion{C}{ii}]$ $\lambda$ 158\ \mbox{$\mu$m}. The highest ionization
lines indicate an \ion{H}{ii} region around WRAY 15-751, while the lowest ionization
lines reveal a photo-dissociation region (PDR). Apart from these emission
lines and the dust continuum, no other dust features have been detected.
It should be noted that Voors et al. (\cite{voo00}) did not detect the
$[\ion{O}{i}]$ $\lambda$ 63\ \mbox{$\mu$m} line on their ISO-LWS
spectrum: only the lines $[\ion{O}{iii}]$ $\lambda$ 88\ \mbox{$\mu$m}
and $[\ion{N}{ii}]$ $\lambda$ 122\ \mbox{$\mu$m} were clearly visible.

\section{Dust continuum emission}
\label{sec:the dust continuum emission}

Aperture photometry was performed on the PACS images and integrated
flux densities derived for the bright nebular shell.
Table~\ref{table:1} presents the Herschel-PACS flux density
measurements, along with data taken from the archives of the Infrared
Astronomical Satellite (IRAS) mission (Neugebauer et
al. \cite{neug84}), the Infrared Space Observatory (ISO) mission
(Kessler et al. \cite{kessler96}) and the Infrared Astronomical
Mission AKARI (Murakami et al. \cite{mura07}). We did not include the
IRAS observation at 100 $\mu$m because it is only an upper limit, and the
AKARI observation at 160 $\mu$m because of its low quality. Note that
the beam size of the IRAS and AKARI observations is large enough to
fully encompass the ring nebula.

Photometric color correction was applied to all flux densities
derived from the data of these four space missions. This correction is
needed to convert monochromatic flux densities that refer to a
constant energy spectrum, to the true object SED flux densities at the
photometric reference wavelengths of each instrument.

\begin{table}[h]
\caption{Color-corrected nebular flux densities.}
\label{table:1}
\centering
\begin{tabular}{l c c c c c}
\hline\hline                            \\
Spacecraft-Instrument&    Date     & $\lambda$  & $F_{\nu}$    & Error  \\
                     &             & ($\mu$m)  & (Jy)    & (Jy)   \\
\hline\hline
IRAS                 &   1983      &   12       &  14.54  & 0.75   \\
                     &             &   25       &  214    & 14  \\
                     &             &   60       &  112    & 12  \\
\hline

ISO-CAM              &  1996       &   10.5     &  8.9    & 0.1    \\
ISO-PHT              &  1996       &   25       &  150    & 40     \\
                     &             &   60       &  75     & 36     \\
                     &             &   105      &  29     & 4      \\
\hline

AKARI-IRC            &  2007       &   9        &  3.32   & 0.03   \\
                     &             &   18       &  82.3  & 2.5   \\ 
AKARI-FIS            &  2007       &   65       &  93.3  & 6.6   \\
                     &             &   90       &  41.4  & 2.7   \\
                     &             &   140      &  15.2  & 1.5   \\
\hline
Herschel-PACS        &  2010       &   70       &  68.9  & 8.3   \\
                     &             &   100      &  31.7  & 5.6   \\
                     &             &   170      &  8.8   & 2.9   \\
\hline\hline
ground-based imaging  &             &            &          &       \\ 
TIMMI-ESO            & 1995        &   10       & 5.6     & 0.1    \\
\hline
\end{tabular}
\end{table}

On the ISO-CAM image\footnote{The 10.5 $\mu$m ISO-CAM image of the
nebula is very similar to the ground-based mid-infrared images
presented in Voors et al. (\cite{voo00}) but its spatial resolution is
much lower.}  the nebular flux density was measured through aperture
photometry, subtracting the contribution from the central object.  For
the color correction of the IRAS data, we used the flux density
ratios to derive the color temperature and then chose the
corresponding color correction factor (Beichman et
al. \cite{beichman}). The ratio R (25,60) corresponds to a temperature
of 190 K, while R (12,25) corresponds to 125 K. We decided to correct
the flux density at 60 $\mu$m using the factor at 190 K. For the flux
densities at 12 and 25 $\mu$m we calculated the corrections using both
the low and the high temperatures and finally considered the
average of the two corrected flux densities, the difference being
accounted for in the errors.  To estimate the color correction of
AKARI FIS and IRC data, we fitted a black body to the two datasets
independently, using the 25 $\mu$m IRAS observation because we needed a
measurement near the maximum of the curve. These fits led us to adopt
the color correction factors that correspond to a temperature of 200
K for FIS (Yamamura et al. \cite{yamamura}) and 150 K for IRC data
(Rosario et al. \cite{rosario}). To color-correct the
Herschel-PACS data, we fitted a black body, considering again the 25
$\mu$m IRAS observation. This fit gave a temperature of 200 K, therefore
we adopted the corresponding correction factor (M\"{u}ller et
al. \cite{muller}).  For the color correction of the ISO data we used
the correction factors given in the corresponding handbooks (Blommaert
et al.  \cite{blom03}; Laureijs et al. \cite{laur03}). Finally, the
mid-infrared flux density derived from ground-based imaging at ESO
with the TIMMI instrument was taken from Voors et al.  (\cite{voo00}).

All these measurements, presented in Table~\ref{table:1}, were considered
to model the dust continuum of the nebula, along with the PACS spectrum,
integrated over the full field of view (25 spaxels) and the archived
ISO-LWS spectrum discussed in Voors et al. (\cite{voo00}).

In Fig.~\ref{fig:sed}, we show the infrared SED of WRAY 15-751 obtained
at different epochs with the various instruments. Within the uncertainties,
all these measurements agree excellently. First, the agreement between
the PACS and ISO-LWS spectra obtained at different epochs, taking into
account that at longer wavelengths the ISO PSF (100\arcsec\ FWHM at
180 $\mu$m) becomes larger than the aperture (84\arcsec) so that some
nebular flux is likely lost, while this is not the case with PACS (PSF
of 14\arcsec\ FWHM at 200 $\mu$m for a 47\arcsec $\times$ 47\arcsec
aperture and a diameter of the nebula smaller than 40\arcsec). Second, the
agreement between the spectra and the photometric data points, indicating
that broad-band photometry is dominated by the dust continuum, and that the
dust shell is well within the PACS spectroscopic field of view.

\subsection{Modeling the dust nebula}

To model and interpret the dust emission spectrum and the
far-infrared images, we used the publicly available two-dimensional
radiative transfer code 2-Dust (Ueta and Meixner \cite{uet03}). 2-Dust
is a versatile code that can be supplied with various grain size
distributions and optical properties as well as complex axisymmetric
density distributions.

\begin{figure}[!]
\resizebox{\hsize}{!}{\includegraphics*{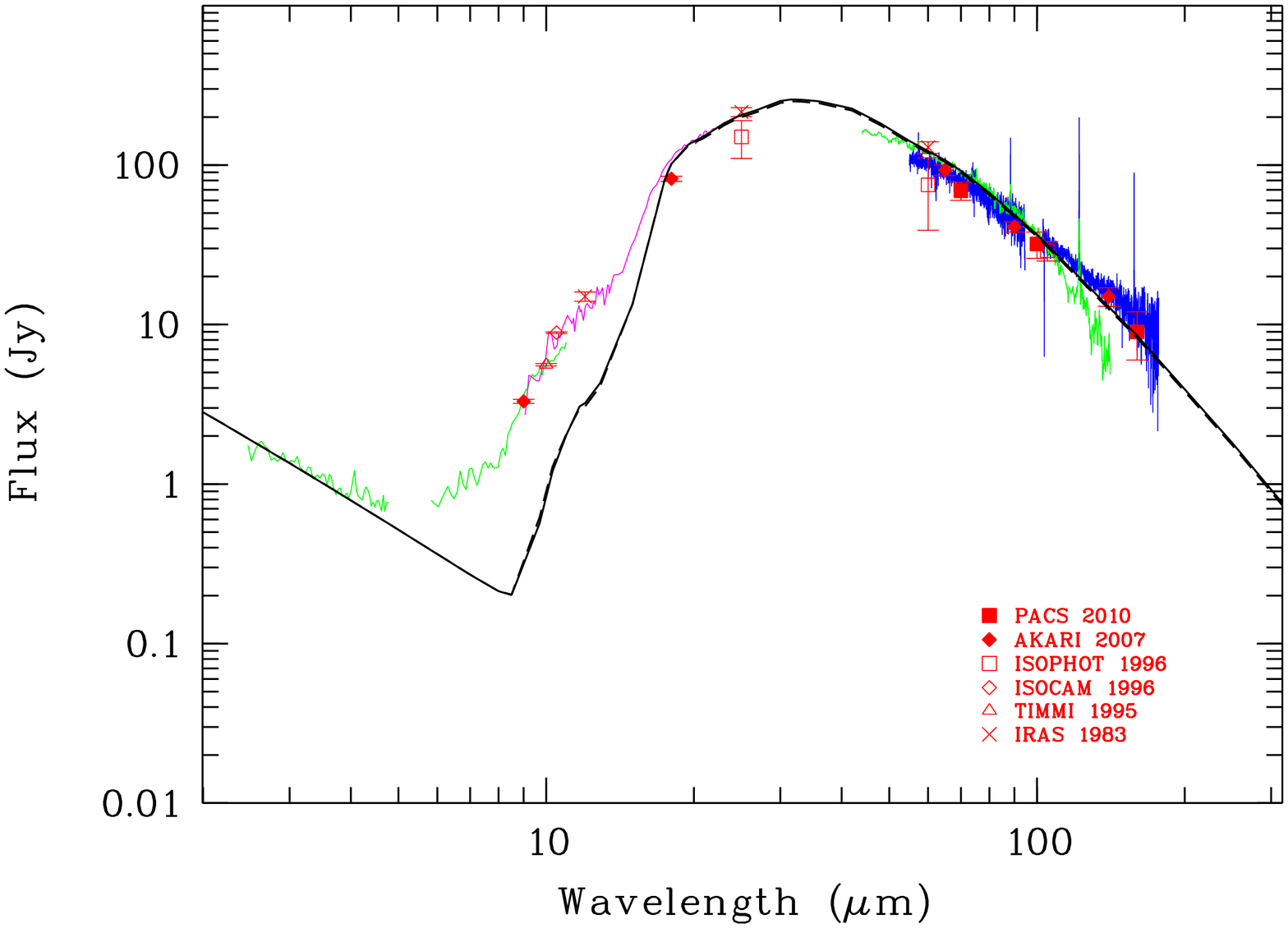}}\\%
\resizebox{\hsize}{!}{\includegraphics*{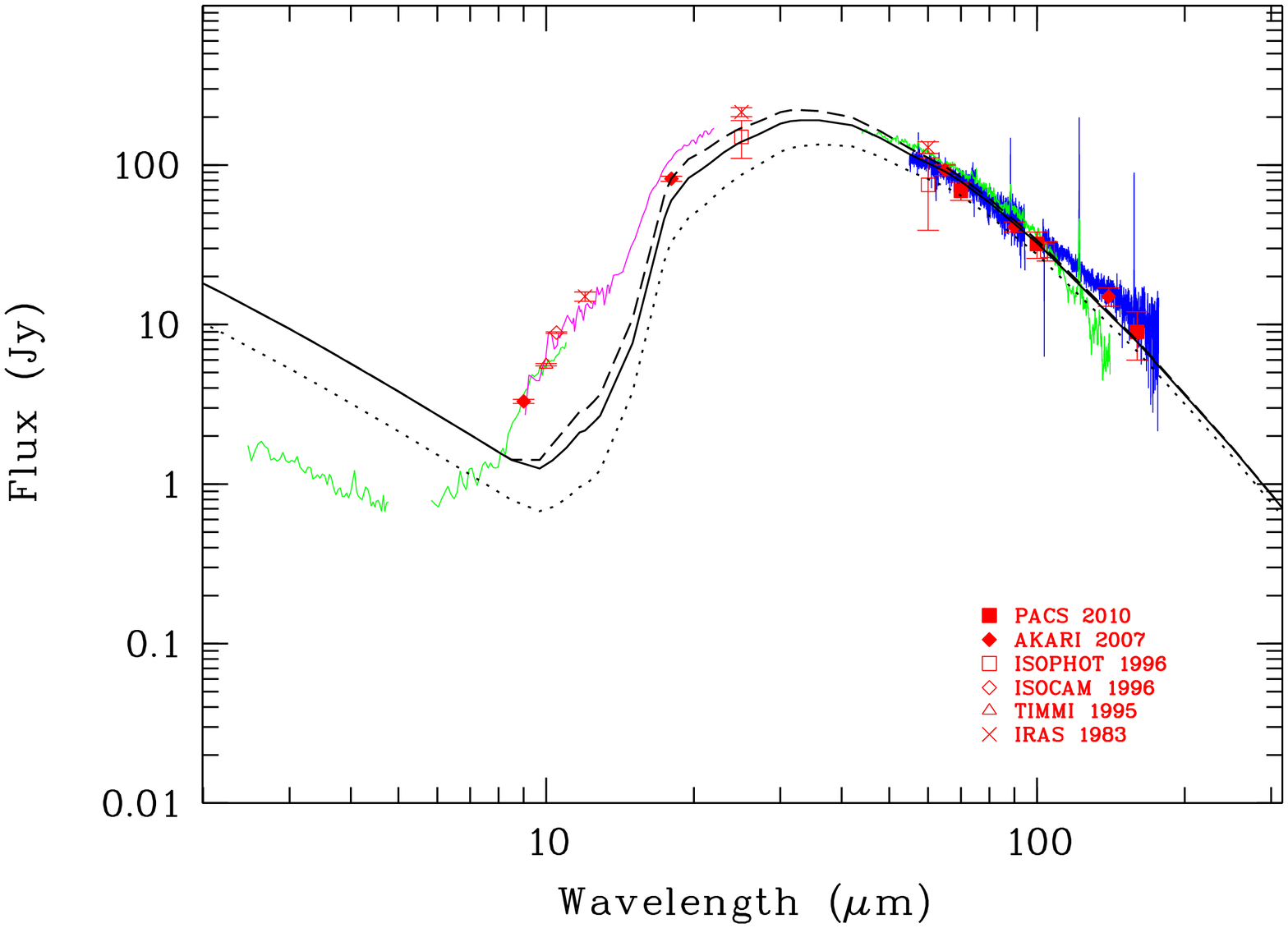}}
\caption{Infrared spectrum of WRAY 15-751 from data collected at
different epochs: IRAS LRS from 8 to 22 $\mu$m (1983, magenta),
ISOPHOT from 2.5 to 11 $\mu$m and ISO LWS from 45 to 140 $\mu$m (1996,
green), Herschel PACS from 55 to 180 $\mu$m (2010, blue). The ISOPHOT
data at $\lambda <$ 5 $\mu$m are corrected for extinction using
$E(B-V)$ = 1.8. At $\lambda >$ 100 $\mu$m, the flux density from ISO LWS is
not reliable, the LWS aperture is too small for the ISO PSF.
Color-corrected photometric measurements are
superimposed (red symbols). The spectrum at $\lambda < $ 5 $\mu$m
comes from the central star, the bump at $\sim$10 $\mu$m from silicates
and very small out-of-equilibrium dust grains, and the spectrum at
$\lambda > $ 20 $\mu$m from the bulk of the dust shell.  Results of
the 2-Dust model fitting are illustrated. Top: best fits of the IRAS/ISO
data assuming $R_{\star}$/$T_{\rm eff}$ = 80/18000, $a_{\rm min}$/$a_{\rm
max}$ = 0.05/1.5 (solid line) and $a_{\rm min}$/$a_{\rm max}$ =
0.2/0.5 (dashed line). Bottom: fits of the AKARI/Herschel data using
the same dust shell properties but with $R_{\star}$/$T_{\rm eff}$ =
320/9000.  Results for a lower-luminosity star $R_{\star}$/$T_{\rm
eff}$ = 240/9000 and $a_{\rm min}$/$a_{\rm max}$ = 0.2/0.5 are also
displayed (dotted line).}
\label{fig:sed}
\end{figure}

Modeling the WRAY 15-751 dust nebula has previously been carried out by
Voors et al. (\cite{voo00}) using IRAS and ISO near- to far-infrared
spectroscopy, mid-infrared (10 $\mu$m) ground-based imaging and a
one-dimensional radiative transfer code. Using the same data and
adopting their input parameters for both the dust and the nebular
properties, we derived quasi-identical results using 2-Dust (e.g.
dust emission spectrum, temperature and mass). In the following,
we use in addition the new PACS imaging and spectroscopic data,
together with AKARI archive data to further constrain the dust
shell properties.

\begin{figure}[!]  \resizebox{\hsize}{!}{%
\includegraphics*[width=5.9cm]{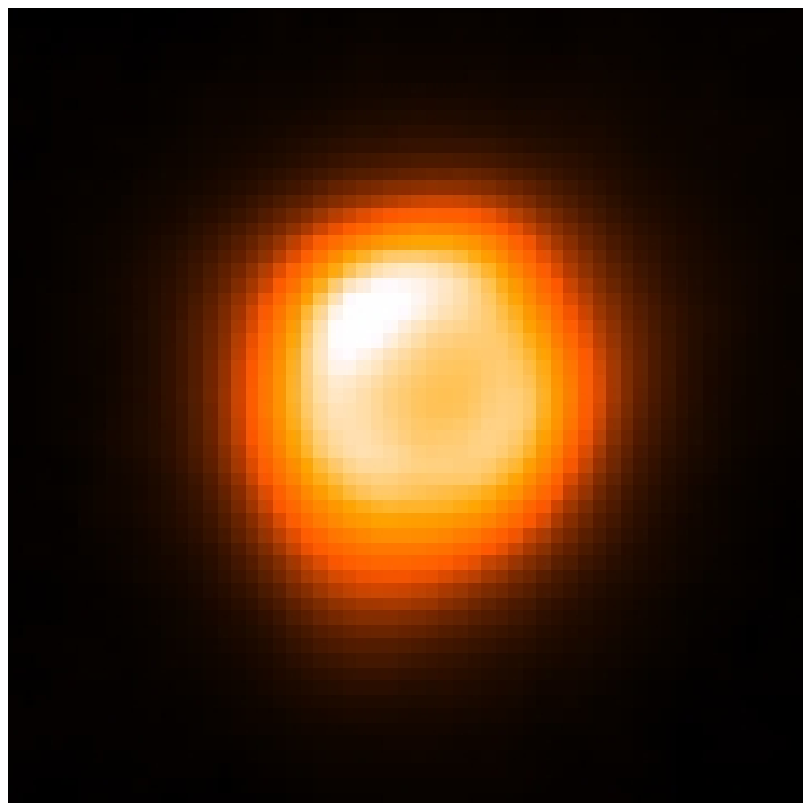}%
\includegraphics*[width=5.9cm]{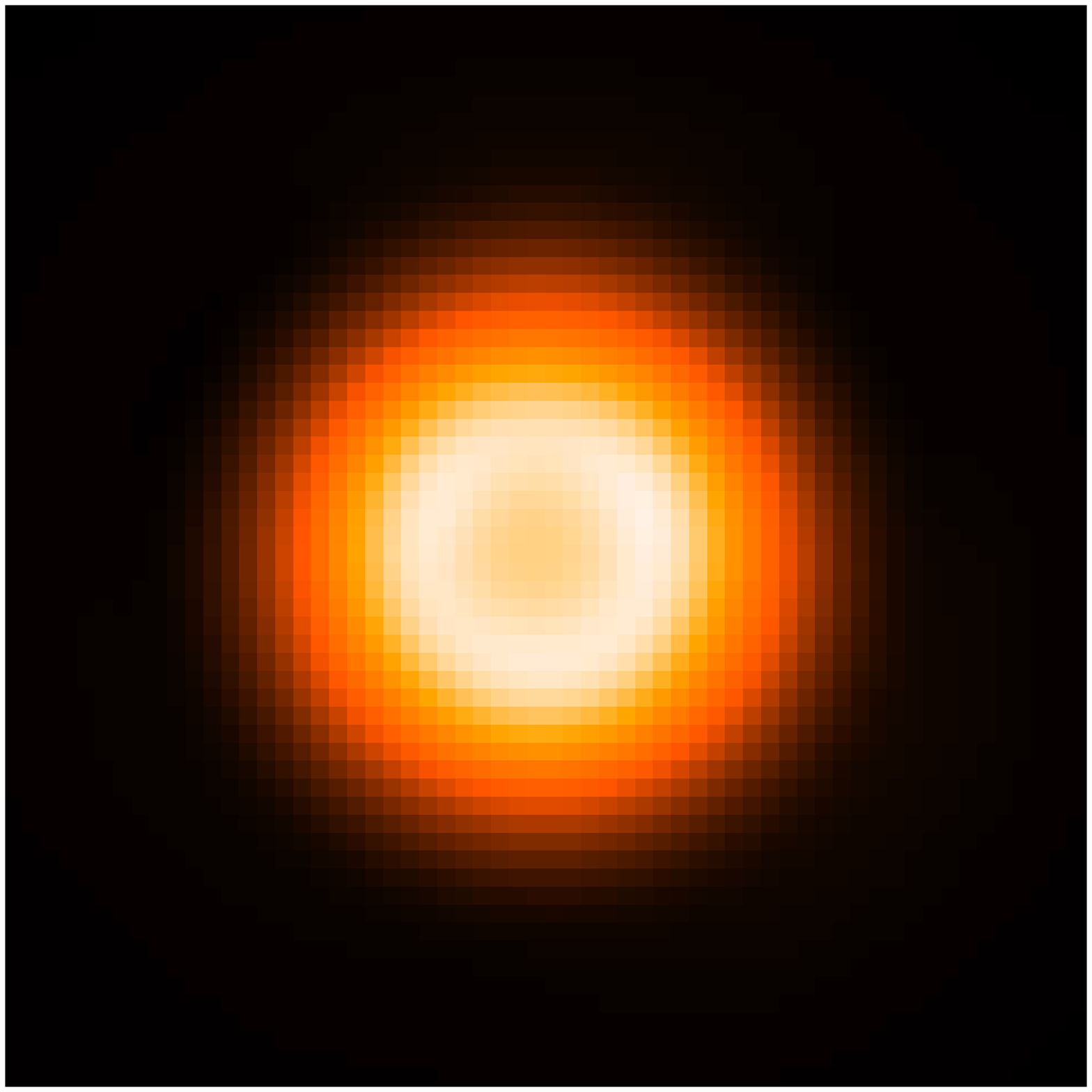}}\\%
\resizebox{\hsize}{!}{\includegraphics*{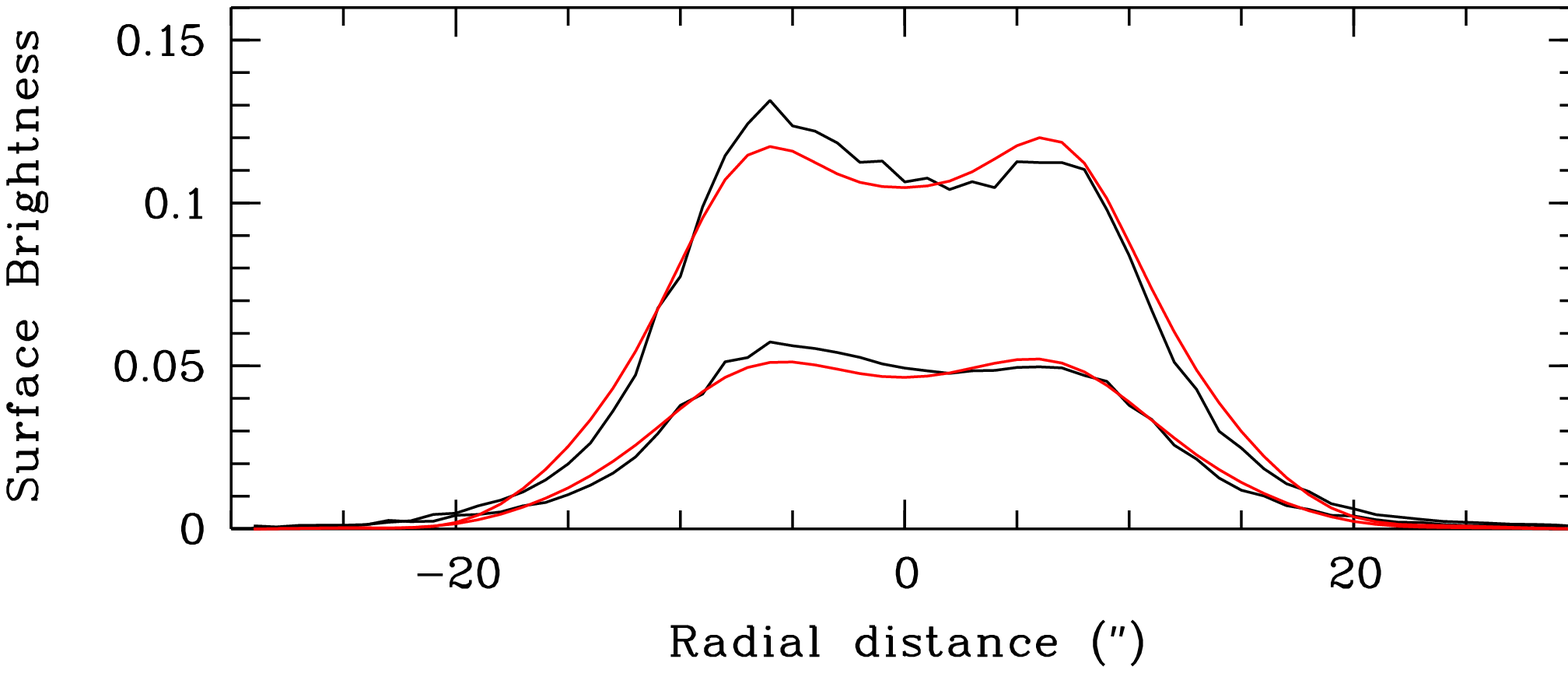}}
\caption{Top left: the $1\arcmin \times 1\arcmin$ image of the nebula
around WRAY 15-751 observed with PACS at 70 $\mu$m. North is up and
east to the left. Top right: the synthetic image computed with 2-Dust
using $r_{\rm in} = 7\arcsec$ and $r_{\rm out} = 21\arcsec$ and
convolved with the PACS PSF.  Bottom: East-west cuts through the
central part of the nebula, observed (black) and synthetic (red). The
upper plots correspond to the 70 $\mu$m image, the lower ones to the
100 $\mu$m image.}
\label{fig:spatcut}
\end{figure}

Voors et al. (\cite{voo00}) showed that the discrepancy between their
model and the data at $\sim$10 $\mu$m (Fig.~\ref{fig:sed}) is probably
caused by a small amount of tiny, warm, out-of-equilibrium carbon grains
in addition to silicates. These warm grains do not significantly
contribute to the bulk of the dust mass at the origin of the emission
at $\lambda >$ 20 $\mu$m. Their mid-infrared image may thus not represent
the main dust component. We then re-derived the inner radius of the dust
shell using the PACS 70 $\mu$m image.  We first assumed that the dust shell
around WRAY 15-751 is spherically symmetric. This is a good proxy to the
overall geometry and limits the number of free parameters.  We also assumed
that the dust density in the nebula runs as $r^{-2}$.  By comparing
the PACS images with the synthetic ones produced by 2-Dust and convolved
with the PACS PSF (Fig.~\ref{fig:spatcut}), we determined the inner
radius of the dust shell, $r_{\rm in} = 7 \arcsec$. This agrees
with the radius derived by Voors et al. (\cite{voo00}). We also adopted
$r_{\rm out} = 3 \times r_{\rm in}$. At a distance of 6 kpc, this
corresponds to $r_{\rm in} =$ 0.20 pc and $r_{\rm out} =$ 0.60 pc.
These results are similar to the measurements discribed in
Sect. 3, considering the errors.

Like other LBVs, WRAY 15-751 exhibits long-term strong photometric
variations.  Since the reaction (heating/cooling) of typical dust
grains to luminosity changes is quasi-instantaneous (e.g.  Bode and
Evans \cite{bode}), the stellar parameters corresponding to
the different epochs of observation must be considered.
In a detailed study, Sterken et al. (\cite{ste08}) showed that
the star was in a minimum (i.e., minimum V brightness), hot
phase in 1989 and in a maximum, cooler phase in 2008. They suggested
that WRAY 15-751 moved in the HR diagram from
$\log L/L_{\odot}$ = 5.9$\pm$0.15, $\log T_{\rm eff}$ =
4.46$\pm$0.02 in 1989 to $\log L/L_{\odot}$ = 5.4$\pm$0.15, $\log
T_{\rm eff}$ = 3.92$\pm$0.02 in 2008. By interpolating, we estimated
$T_{\rm eff}$ = 18000~K in 1996, at the epoch of the ISO observations.
A good fit of the ISOPHOT stellar spectrum at $\lambda <$ 5 $\mu$m is
obtained with $R_{\star}$ = 80 $R_{\odot}$, which corresponds to a
stellar luminosity $\log L/L_{\odot}$ = 5.8. Unfortunately, the
photometric measurements are very scarce before 1989. From the V light
curve displayed by Sterken et al.  (\cite{ste08}), the brightness of
the star seems nevertheless similar in 1983, the epoch of the IRAS
observations.  We then adopted $T_{\rm eff}$ = 18000~K and $R_{\star}$ =
80 $R_{\odot}$ as input for the 2-Dust modeling of both the IRAS and
ISO data sets. At the epoch of the AKARI and Herschel
observations in 2007-2010, the star is much cooler and apparently less
luminous. This is quite surprising given the good agreement of the
IRAS/ISO and the AKARI/Herschel spectroscopic and photometric data
seen in Fig.~\ref{fig:sed}. Although AKARI/Herschel flux densities might be
marginally lower than the IRAS/ISO ones, this constitutes a strong
constraint for the modeling since the nebula itself cannot have
significantly changed between 1996 and 2007-2010. To model the
2007-2010 data, we considered two sets of stellar parameters: a constant-
luminosity, low-temperature model with $T_{\rm eff}$ = 9000~K,
$R_{\star}$ = 320 $R_{\odot}$, and a low-luminosity, low-temperature
one with $T_{\rm eff}$ = 9000~K, $R_{\star}$ = 240 $R_{\odot}$ (i.e.,
$\log L/L_{\odot}$ = 5.5) which better agrees with the most recent position
of WRAY 15-751 in the HR diagram estimated by Sterken et
al. (\cite{ste08}). Note that we neglected the delay in the
response of the different parts of the dust shell to stellar changes,
at most about four years in the observer frame for a shell radius of 0.6 pc.
Such a delay will mostly smear out the effects of the stellar variations over
some years.

Voors (\cite{voo99}) and Voors et al. (\cite{voo00}) found that the
dust in the WRAY 15-751 nebula is dominated by amorphous silicates,
with little contribution from crystalline species. They also obtained
a best fit of the spectrum using pyroxenes and a 50/50 Fe to Mg
abundance. We therefore started with a similar dust composition, using
the optical constants given by Dorschner et al. (\cite{dor95}),
extrapolated to a constant refraction index in the far-ultraviolet.
We assumed the size distribution for the dust grains of Mathis et
al. (\cite{mat77}, hereafter MRN): $n(a) \propto a^{-3.5}$ with $a_{\rm min} < a <
a_{\rm max}$, $a$ denoting the grain radius. By varying the opacity,
which controls the strength of the emission, and $a_{\rm max}$ (or
$a_{\rm min}$), which controls the 20$\mu$m / 100$\mu$m flux density ratio,
several good fits can be obtained (we did not attempt to fit the 10
$\mu$m bump, which is due to out-of-equilibrium dust, only a
minor contributor to the dust mass). Acceptable values of $a_{\rm
max}$ range between 0.5 and 1.5 $\mu$m, confirming the presence of
large $\sim$1 $\mu$m dust grains in the nebula. In all cases the
nebula is optically thin, the opacity is lower than 0.01 at 25
$\mu$m.

However, when a good fit of the IRAS/ISO data was obtained with the hot
$R_{\star}$/$T_{\rm eff}$ = 80/18000 stellar parameters, we were unable
to reproduce the AKARI/Herschel data using the same dust shell
properties with the cooler star, even when using the constant
luminosity 320/9000 model. A higher luminosity central star would be
needed to compensate for the shift of stellar energy output from lower to
higher wavelengths. We then tried to increase the near-infrared dust
absorptivity by increasing the Fe to Mg ratio. Using the optical data
of pyroxenes with a higher Fe to Mg ratio provided by Dorschner et
al. (\cite{dor95}), the fit can be improved, but not sufficiently so. We
then used the silicate dust with the highest available near-infrared
absorptivity, i.e., the optical data given by Ossenkopf et
al. (\cite{oss92}) for cold O-rich silicate with Fe inclusions (see
also Fig.~7 of Dorschner et al. \cite{dor95}), with an average
bulk density $\rho$ = 3.5 g cm$^{-3}$. Using a
narrow range of dust radii, $a_{\rm min}$ = 0.2 $< a <$ $a_{\rm max}$
= 0.5, we were finally able to fit both the IRAS/ISO data with the
$R_{\star}$/$T_{\rm eff}$ = 80/18000 model and the AKARI/Herschel data
with the $R_{\star}$/$T_{\rm eff}$ = 320/9000 model
(Fig.~\ref{fig:sed}). The observed dust emission cannot be
reproduced when using the low-luminosity stellar parameters
$R_{\star}$/$T_{\rm eff}$ = 240/9000 suggested by Sterken et al.
(\cite{ste08}) for the 2007-2010 epoch.

In summary, the nebular dust emission can be reproduced for both the
hot and cool stellar phases, assuming a constant stellar luminosity
and Fe-rich dust grains. The adopted range of grain radii is
unrealistically narrow, but numerical tests show that wider ranges can
be considered if the near-infrared absorptivity is increased even
more. As supported by the modeling, the fact that the dust
emission does not significantly change from 1996 to 2010 suggests that
stellar variations occur at essentially constant luminosity, as found
in several LBVs (Wolf et al. \cite{wol81}, Humphreys and Davidson
\cite{hum94}). Our results are not entirely incompatible with those of
Sterken et al. (\cite{ste08}) since the uncertainties on the stellar
luminosity are large and possibly underestimated, as quoted by the
authors themselves.  Our results demonstrate that the far-infrared
dust emission from dust shells can be used to constrain the luminosity
variations of the central star.

The mass and temperature of the dust shell we derive from the modeling
depend little on the exact stellar parameters and dust sizes, provided
that a good fit of the spectrum is obtained. We find that the total
dust mass in the nebula is $M_{\rm dust} = 4.5 \pm 0.5 \times 10^{-2}$
M$_{\odot}$ and that $T_{\rm dust}$ varies from 95 K at $r_{\rm in}$
to 66 K at $r_{\rm out}$. The quoted uncertainty of $M_{\rm dust}$
only accounts for the dispersion of the values obtained with different
models and is therefore underestimated.  Our value of $M_{\rm dust}$
is higher than the one derived by Voors et al. (\cite{voo00}) mainly
because we used a larger distance to WRAY 15-751.

It is interesting to compare these estimates with those determined using
empirical methods.  Indeed, $M_{\rm dust}$ can be derived using
\begin{equation} M_{\rm dust} = \frac{F_{\nu} \, \, D^{2}}{B_{\nu}
\left( T_{\rm dust} \right) \, K_{\nu}} \;,
\end{equation} where $K_{\nu}$ is the mass absorption coefficient,
$B_{\nu}$ the Planck function and $D$ the distance to the nebula
(Hildebrand \cite{hil83}). $K_{\nu}$ is roughly independent of the
grain radius and behaves as $\nu^{\beta}$ in the far-infrared. For the
cold O-rich silicates of Ossenkopf et al. (\cite{oss92}), $K_{60}$ =
50 cm$^{2}$g$^{-1}$ at 60 $\mu$m and $\beta$ = 2.1. By fitting a
modified black-body with $\beta$ = 2.1 to the $\lambda \geq$ 18 $\mu$m
photometric data points (Fig.~\ref{fig:sed}), we obtain $T_{\rm dust}$
= 87 K, not far from the mean value of the temperatures found with
2-Dust. The higher ${T_{\rm dust}}$ and lower $\beta$ obtained in
Vamvatira-Nakou et al. (\cite{vam11}) are due to the inclusion in the
fit of the data at $\sim$ 10 $\mu$m. Using Eq.~1 and the
color-corrected IRAS flux density $F_{60}$ = 129 Jy, we obtain $M_{\rm
dust} \simeq 3.5 \times 10^{-2}$ M$_{\odot}$, which agrees reasonably
well with the value derived with 2-Dust, given the large
uncertainties. The lower values of $M_{\rm dust}$ reported in
Hutsem\'ekers (\cite{hut94,hut97}) using the same method are
essentially due to the use of different values of $K_{\nu}$ and
$\beta$.

\subsection{Properties of the outer nebula}

We now estimate the temperature and the mass of the outer nebula (shown
in Fig.~\ref{2shell}).

After carefully subtracting the background and the bright inner
dust shell, we measured $F_{100}$ = 5.0$\pm$0.5 Jy at 100 $\mu$m where
the outer nebula is best seen, and $F_{70}$ = 6.0$\pm$0.5 Jy at 70
$\mu$m. At 160 $\mu$m, we estimated $F_{160}$ = 5.0$\pm$1.5 Jy, but
this value strongly depends on the reduction procedure and background
subtraction.

Assuming the same dust composition for the outer shell as for the
inner one, i.e., $\beta \simeq 2$ (silicates), we derived ${T_{\rm
dust}}$ = 40$\pm$5~K by fitting a modified black-body to the flux densities
measured at 70 and 100 $\mu$m.  Within the uncertainties, the flux density at
160 $\mu$m is barely compatible with this temperature, most likely due to
background contamination. The temperature of the nebula is higher
than the temperature of the nearby background emission measured around ${T_{\rm
dust}}$ $\simeq$ 20 K, thus supporting the association of the outer nebula with WRAY
15-751. It is also interesting to note that for silicates,
${T_{\rm dust}}$ is expected to vary as as $r^{-1/3}$ (e.g. Tielens
\cite{tie05}), so that the second shell of radius $\sim$ 70\arcsec\
should have ${T_{\rm dust}} \simeq$ 45~K extrapolating from the average 
temperature of the inner shell. This is consistent with the measured value.

Using $F_{100} \simeq$ 5.0$\pm$0.5 Jy, $K_{100}$ = 18 cm$^{2}$g$^{-1}$
and ${T_{\rm dust}}$ = 40$\pm$5 K in Eq.~1, we find $M_{\rm dust}$ =
$5 \pm 2 \times 10^{-2}$ M$_{\odot}$.  Although uncertain, the mass of
dust in the outer shell appears at least as large as the mass in the
bright inner shell.

\section{Emission line spectrum}
\label{sec:the analysis of the emission line spectrum}

\subsection{Line flux measurements}

\begin{table}[t]
\caption{Line fluxes from the summed spectrum} 
\label{table:2}
\centering
\begin{tabular}{l c c c c}
\hline\hline
Ion & $\lambda$ & Band & $F\ $ (9 spaxels) & $ F\ $ (corrected)  \\
    &  ($\mu$m)    &   & ($10^{-15}$ W~m$^{-2}$)          & ($10^{-15}$ W~m$^{-2}$) \\
\hline\hline
    $[\ion{N}{iii}]$ & 57  & B2A   & 1.60 $\pm$ 0.38 &  \\
                     &     & B3A   & 1.36 $\pm$ 0.30 &  \\
                     &     & Mean  & 1.45 $\pm$ 0.24 &  1.73 $\pm$ 0.29\\
    $[\ion{O}{i}]$   & 63  & B2A   & 1.06 $\pm$ 0.26 &  \\
                     &     & B3A   & 0.93 $\pm$ 0.21 &  \\
                     &     & Mean  & 0.98 $\pm$ 0.16 &  1.18 $\pm$ 0.19\\
    $[\ion{O}{iii}]$ & 88  & B2B   & 0.83 $\pm$ 0.18 &  1.04 $\pm$ 0.23\\
    $[\ion{N}{ii}]$  & 122 & R1B   & 4.14 $\pm$ 0.82 &  5.45 $\pm$ 1.08\\
    $[\ion{O}{i}]$   & 146 & R1B   & 0.10 $\pm$ 0.03 &  0.14 $\pm$ 0.04\\
    $[\ion{C}{ii}]$  & 158 & R1B   & 0.81 $\pm$ 0.16 &  \\
                     &     & R1A   & 0.96 $\pm$ 0.19 &  \\
                     &     & Mean  & 0.87 $\pm$ 0.12 &  1.21 $\pm$ 0.17\\
    $[\ion{N}{ii}]$  & 205 & R1A   & 0.97 $\pm$ 0.24\tablefootmark{a} &  1.47 $\pm$ 0.36\\
\hline
\end{tabular} \\
\tablefoottext{a}{Corrected value from PACS/SPIRE cross-calibration}
\end{table}

We measured the emission line intensities in each one of the 25
spectra (Fig.~\ref{wra_foot}) by fitting a Gaussian to the line
profiles using IRAF.  The detailed measurements are given in Appendix
A. Only at the central spaxel (2,2) are all the lines detected. At
this spaxel the intensities of almost all lines reach their highest
values. Most of the flux is detected in the $3\times3$ spaxels central
area. In contrast, the outer 16 spaxels contribute significantly less
to the line fluxes.

To investigate whether there are differences in the properties of
the gas in different parts of the nebula, we also computed for each
spaxel the flux ratios of every detected line to the
line $[\ion{N}{ii}]$ $\lambda$ 122\ \mbox{$\mu$m}, which is the
strongest one. There is some evidence that the ratio $[\ion{C}{ii}]$
158 $\mu$ m / $[\ion{N}{ii}]$ 122 $\mu$m might be higher in the outer
spaxels than in the central ones. However, this trend is not
significant given the large errors.  Consequently, we cannot conclude
that there is any clear trend with the distance to the center.

To measure the total emission line fluxes in the nebula with a
reasonable accuracy, in particular to compute diagnostic flux ratios,
we used the sum of the spectra that correspond to the 9 ($3\times3$)
central spaxels. The 16 outer spaxels are not included in the sum because
they bring more noise than signal especially for the faint lines. We
again calculated the line flux by fitting a Gaussian
profile to each one of the detected forbidden emission lines. The
results are given in Table~\ref{table:2}.  This table contains the
detected ions, the wavelength of each line, the corresponding
spectral waveband in which they were detected, and the fluxes with
their errors.  The quoted uncertainties are the sum of the line
fitting uncertainty plus the uncertainty due to the position of the
continuum, to which we quadratically added an error of 20\% to account
for the uncertainty of the PACS absolute flux calibration. Note that
within a given waveband, relative flux uncertainties are smaller, on
the order of 10\%. There is a good agreement between the fluxes
measured in two different bands for a given emission line so that
weighted mean values are computed. The line $[\ion{N}{ii}]$ $\lambda$205\
\mbox{$\mu$m} had a problematic calibration in PACS. Therefore, to
be able to use the corresponding flux values for the following analysis,
we calculated a correction factor using objects from the MESS
collaboration (Groenewegen et al. \cite{groenewegen}) observed with
both PACS and SPIRE.  Then, from the SPIRE/PACS cross calibration we
found that the measured $[\ion{N}{ii}]$ $\lambda$205\ \mbox{$\mu$m}
flux should be multiplied by a correction factor of 5.5. The error of
the final corrected $[\ion{N}{ii}]$ 205\ \mbox{$\mu$m} fluxes is
assumed to be 25\%.

When using the central $3\times3$ spaxel region, some nebular flux is lost,
the amount of which depends on the wavelength as the beam size, with
consequences on the flux ratios. On the other hand, the spectrum summed
over all 25 spaxels encompasses the full ring nebula, as shown in
Fig.~\ref{wra_foot} and supported by the agreement with the photometric
measurements (Sect. \ref{sec:the dust continuum emission}). Thus, assuming
that the spectral lines originate from the same regions as the dust continuum,
we used the ratio of the 9-spaxel continuum spectrum to the 25-spaxel
continuum spectrum to estimate the correction factor, which varies
roughly linearly from 0.85 at 50 $\mu$m to 0.65 at 210
$\mu$m. Corrected flux values are given in the rightmost column of
Table~\ref{table:2}. For the two lines reasonably detected outside the
central area, i.e., $[\ion{N}{ii}]$ $\lambda$122\ \mbox{$\mu$m} and
$[\ion{C}{ii}]$ $\lambda$158\ \mbox{$\mu$m} (Table \ref{table:5}), we
directly measured the fluxes integrated over the 25 spaxels. We found
$F$ = 5.0 $\pm$ 1.0 $\times$ $10^{-15}$ W~m$^{-2}$ for $[\ion{N}{ii}]$
and $F$ = 1.31 $\pm$ 0.18 $\times$ $10^{-15}$ W~m$^{-2}$ for
$[\ion{C}{ii}]$, in good agreement with the corrected values given in
Table~\ref{table:2}.

\subsection{Photoionization region characteristics}

The emission lines associated to the \ion{H}{ii} region detected in
the spectrum of the inner nebula are $[\ion{N}{iii}]$ 57\
\mbox{$\mu$m}, $[\ion{O}{iii}]$ 88\ \mbox{$\mu$m}, and $[\ion{N}{ii}]$
122, 205\ \mbox{$\mu$m}. The other three emission lines originate from
a region of transition between ionized and neutral hydrogen,
indicating a photodissociation region (PDR). Extensive analysis and
discussion of the latter lines is given in the next section.

\subsubsection{H$\alpha$ flux}

The H$\alpha$ flux from the nebula was estimated by integrating the
surface brightness over the whole nebula
(Fig.~\ref{Halpha}). Contamination by field stars was corrected for,
and emission from the central part extrapolated using the mean surface
brightness.  The contribution of the strong $[\ion{N}{ii}]$ lines was
removed using the $[\ion{N}{ii}]$ /H$\alpha$ ratio measured in HVD and
the transmission curve of the H$\alpha$+$[\ion{N}{ii}]$ filter. The
conversion to absolute flux was made with the help of
spectrophotometric standard stars observed in the same filter.
Adopting a color excess E(B$-$V)=1.8 $\pm$ 0.3 based on the available
optical studies of the nebula (Hu et al. \cite{hu}; HVD \cite{hut91};
Voors et al.  \cite{voo00}; Garcia-Lario et al. \cite{garcia}), we
finally derived $F_{0}(\mathrm{H}\alpha$) = 3.1 $\times$ 10$^{-11}$
ergs~cm$^{-2}$~s$^{-1}$ (=3.1 $\times$ 10$^{-14}$ W~m$^{-2}$). The
uncertainty of this value amounts to $\sim$20\%. It is more accurate
than --and agrees with-- the value given by Hutsem\'ekers
(\cite{hut94}).

\subsubsection{Electron density}

The $[\ion{N}{ii}]$ 122/205\ \mbox{$\mu$m} ratio, equal to 3.71 $\pm$
1.17, provides a diagnostics for the electron density,
$n_\mathrm{e}$. To calculate it we used the package \textit{nebular}
of the IRAF/STSDAS environment (Shaw \& Dufour \cite{shaw}). This
algorithm makes use of the fact that the nebular cooling-rate is
dominated by ions, most of which have either $p^2$, $p^3$ or $p^4$
ground-state electron configurations. These configurations have five
low-lying levels. The main physical assumption is that only these
five levels are considered to calculate the emission line
spectrum.  For all the following calculations, an electron temperature
constant throughout the nebula and equal to $T_\mathrm{e}=10^4\
\mathrm{K}$ was assumed with an uncertainty of 20\%.  This value
is reasonable since we observe higher excitation (i.e., the $[\ion{N}
{iii}]$ 57\ \mbox{$\mu$m} and $[\ion{O}{iii}]$ 88\
\mbox{$\mu$m} lines) than in the AG Car nebula, for which Smith et
al. (\cite{smith97_2}) calculated an electron temperature between 5900
to 7000 K.  The electron density, using the $[\ion{N}{ii}]$ 122/205\
\mbox{$\mu$m} ratio, is found to be $164 \pm 90\ \mathrm{cm}^{-3}$.

The $[\ion{S}{ii}]$ 6716/6731\ \AA\ ratio is also an electron density
diagnostics. The value of this ratio measured by HVD is equal to
1.1$\pm$0.1, which yields to an electron density of $423 \pm 183
\ \mathrm{cm}^{-3}$, using the same tool and hypothesis. For
the following analysis, we used the average electron density,
i.e. $n_\mathrm{e}=210 \pm 80\ \mathrm{cm}^{-3}$, a typical value
for LBV nebulae (Nota et al. \cite{nota1}).

\subsubsection{Ionizing flux}

It should be noted here that the recombination time in our case is
much longer than the timescale of the variability exhibited by the
central star of the nebula. More precisely, the recombination time is
equal to $\tau_{rec}= 1/n_\mathrm{e}\alpha_\mathrm{B}=(1.22 \times
10^{5}/n_\mathrm{e})\ \mathrm{yr}$ (Draine \cite{draine11}), where
$\alpha_\mathrm{B}$ is the recombination coefficient. Using the
measured electron density, we estimated that the recombination time is
about 440 yr. Consequently, the stellar variations of $\sim$ 10 yr
cannot change the photoionization/recombination timescale
significantly and an average nonvariable star can be considered.

The rate of emission of hydrogen-ionizing photons, $Q_{0}$, and the
Str\"omgren radius of the ionized hydrogen region, $R_{S}$, can thus
be determined. The nebula was considered to be spherical with an
uniform density. $Q_{0}$ and $R_{S}$ were first determined using the
estimated H$\alpha$ flux and second based on the radio flux density,
$S_\nu$ = 24 mJy at 6 cm (4.9 GHz) which was taken from the study of
Duncan and White (\cite{duncan}), adopting a typical error of 0.5
mJy. It should be mentioned here that the nebula is optically thin at
4.9 GHz, as the optical depth, calculated using equation (B.15), is
lower than one.

The following equation gives the $R_{S}$ in pc (see Appendix B)
       \begin{equation}
        R_{\mathrm{S}}=3.17\left(\frac{x_e}{\epsilon}\right)^{1/3}
                      \left(\frac{n_\mathrm{e}}{100}\right)^{-2/3}
                       T_4^{(0.272+0.007\mathrm{ln}T_4)}\left
                       (\frac{Q_0}{10^{49}}\right)^{1/3},
       \end{equation}
where, using the H$\alpha$ flux, $Q_{0}$ (in photons $\mathrm{s}^{-1}$) is given by
       \begin{equation}
        Q_{0(\mathrm{H\alpha})}=8.59\times10^{55}
                  T_4^{(0.126+0.01\mathrm{ln}T_4)}D^2F_0(\mathrm{H}_{\alpha}) \; .
       \end{equation}
When using the radio flux, $Q_{0}$ (in photons $\mathrm{s}^{-1}$)
is given by
       \begin{equation}
        Q_{0(\mathrm{radio})}=8.72\times10^{43}T_4^
                     {(-0.466-0.0208\mathrm{ln}T_4)}\left(\frac{\nu}{4.9}
                     \right)^{0.1}x_e^{-1}D^2S_{\nu} \; .
       \end{equation}
In these equations $x_e=n_e/n_p$, i.e. the fraction of the electron
density to the proton density, $\epsilon$ is the filling factor,
$T_4=T_e/(10^4\ \mathrm{K})$, $\nu$ is the radio frequency (4.9 GHz in
this case) and $\textit{D}$ is the distance of the nebula in kpc. The
H$\alpha$ flux, $F_0(H\alpha)$, is in ergs~cm$^{-2}$~s$^{-1}$, while
the radio flux, $S_{\nu}$, is in mJy.

Using the above equations and assuming $x_e=1$ (the star is not hot
enough to significantly ionize He), $\epsilon=1$ (the whole volume of
the nebula is filled by ionized gas) and $T_4=1$, the rate of emission
of hydrogen-ionizing photons is found to be
$Q_{0(\mathrm{H\alpha})}=(9.6\pm3.7) \times
10^{46}~\mathrm{photons~s^{-1}}$ and
$Q_{0(\mathrm{radio})}=(7.5\pm2.5) \times 10^{46}\
\mathrm{photons~s^{-1}}$.  Within the uncertainties, these two results
agree well. This also means that the adopted
value of E(B$-$V) is essentially correct. The mean value is $Q_{0}=
(8.2\pm2.1)\times10^{46}\ \mathrm{photons~s^{-1}}$ and corresponds to
an early-B star, $T_{\mathrm{eff}}\sim22000\ \mathrm{K}$ (Panagia
\cite{pan73}), in agreement with the average spectral type of the
star (Hu et al. \cite{hu}; Sterken et al. \cite{ste08}).

\begin{figure}[t]
\resizebox{\hsize}{!}{\includegraphics{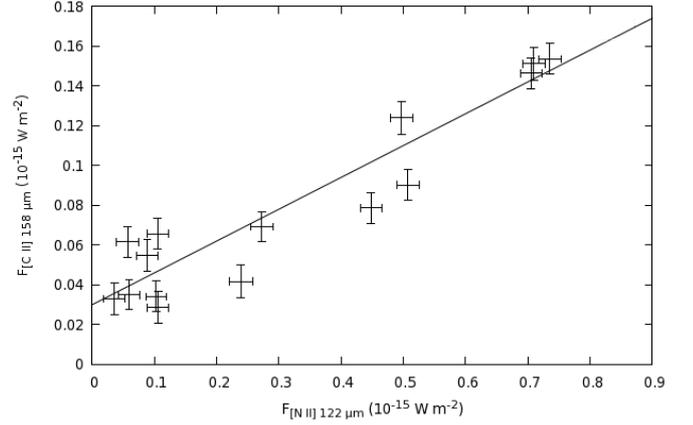}}
\caption{Correlation between the $[\ion{C}{ii}]$ 158\ \mbox{$\mu$m}
         and $[\ion{N}{ii}]$ 122\ \mbox{$\mu$m} line fluxes for each
         spaxel where these two lines are detected.}
\label{pdr_correlation}
\end{figure} 

The Str\"omgren radius calculated from equation (2) is
$R_{S}=0.46\pm0.17$ pc.  By definition, the Str\"omgren radius is the
radius of an ionization bounded nebula.  In section~\ref{sec:morphology
of the nebula} the radius of the nebula in the optical, which is the
radius of the ionized gas region which surrounds the central star, was
found to be 0.32 pc. As both radii agree within the errors, we can
conclude that the nebula can be ionization bounded, in agreement with
the presence of PDR lines in the spectrum.

\subsubsection{Abundances}

The N/O abundance ratio can be estimated using the lines
$[\ion{N}{iii}]$ 57\ \mbox{$\mu$m} and $[\ion{O}{iii}]$ 88\ \mbox{$\mu$m} and
the equation
      \begin{equation}
          \frac{\mathrm{N}}{\mathrm{O}}=\frac{\langle \mathrm{N}^{++}\rangle}
                        {\langle \mathrm{O}^{++}\rangle}
                       =\frac{F_{[\ion{N}{iii}]57}/\varepsilon_{[\ion{N}{iii}]57}}
                             {F_{[\ion{O}{iii}]88}/\varepsilon_{[\ion{O}{iii}]88}} \,,
      \end{equation}
where \textit{F} is the observed line flux and $\varepsilon$ is the
volume emissivity. Considering $T_\mathrm{e}=10^4\ \mathrm{K}$ and
$n_\mathrm{e}=210\ \mathrm{cm}^{-3}$, we derived the emissivities
using the package \textit{nebular}. From the measured line intensities
(Table~\ref{table:2}), the N/O abundance ratio is then found to be 1.00
$\pm$ 0.38. The N/O ratio is much higher than the solar value
of 0.14 (Grevesse et al. \cite{grevesse}). Compared with the N/O ratios
of other LBVs nebulae (Smith \cite{smith97}; Smith et
al. \cite{smith98}; Lamers et al. \cite{lamers}), the value in the WRAY
15-751 nebula is one of the lowest and almost the same as the value
found for the LMC R127 nebula, 0.9 $\pm$ 0.4.

An estimate of the N/H abundance number ratio can also be made, based
on the observed H$\alpha$ 6562.8 \AA, $[\ion{N}{iii}]$ 57\
\mbox{$\mu$m}, $[\ion{N}{ii}]$ 122\ \mbox{$\mu$m} and 205\
\mbox{$\mu$m} lines, considering that
      \begin{equation}
          \frac{\mathrm{N}}{\mathrm{H}}=\frac{\langle \mathrm{N}^{+}\rangle
             +\langle \mathrm{N}^{++}\rangle}{\langle \mathrm{H}^{+}\rangle}\ ,
      \end{equation}
The flux ratios, $F/F_0 (\mathrm{H}\beta)$ were calculated for the
three infrared lines of nitrogen. The observed values of $\textit{F}$
were taken from Table~\ref{table:2}.  To calculate the H$\beta$ flux,
given the dereddened H$\alpha$ flux, we assumed a case-B recombination
with $T_\mathrm{e}=10^4 $ K. The ionic abundances $\mathrm{N}^{+}
/\mathrm{H}^{+}$ and $\mathrm{N}^{++}/\mathrm{H}^{+}$ where then
derived using again the package \textit{nebular}. Their sum gives the
N/H abundance number ratio, calculated to be $(4.3\pm2.0)
\times10^{-4}$. This value is equivalent to a logarithmic N/H
abundance of 12 + log(N/H) = 8.63 $\pm$ 0.20, higher than the solar
value of 7.83 (Grevesse et al. \cite{grevesse}). This value is similar
to the N/H abundances of other LBV nebulae (Smith \cite{smith97};
Smith et al. \cite{smith98}; Lamers et al. \cite{lamers}).

\subsubsection{Mass of the ionized gas}

The mass of the ionized gas can be estimated based on the H$\alpha$
and the radio emissions. For this calculation the equations derived 
in Appendix B were used.

Since the temperature of the central star is lower than 30000 K, we
can assume that the ionization of He is negligible ($y_{+} =
0$). Assuming also $\epsilon=1$, the mass of the ionized nebula is
$M_{i(\mathrm{H\alpha})} =1.04\pm0.53\ \mathrm{M}_{\odot}$ and
$M_{i(\mathrm{radio})}=0.92\pm0.46\ \mathrm{M}_{\odot}$. The average
value is $M_{i} =0.97\pm0.35\ \mathrm{M}_{\odot}$.  If the
nebula is considered to be a shell and not a sphere, with inner
radius 7\arcsec\ (assuming that the H$\alpha$ shell has the same
inner radius as the infrared dust shell) and outer radius 11\arcsec\ ,
which is the limit of the  H$\alpha$ nebula as described in Sect. 3,
its ionized mass is $M_{i}=0.84\pm0.31\ \mathrm{M}_{\odot}$. Considering
the errors, the ionized mass in the case of a shell nebula is not
significantly different from the spherical case.

\subsection{Photodissociation region characteristics}

The fine structure lines $[\ion{O}{i}]$ 63, 146\ \mbox{$\mu$m} and
$[\ion{C}{ii}]$ 158\ \mbox{$\mu$m} indicate a PDR in
the nebula because they are among the important coolants in PDRs
(Hollenbach \& Tielens \cite{hollenbach}). In this region, which
surrounds the ionized region of the nebula, the gas is neutral and the
far-ultraviolet (FUV) photons (with $h\nu<13.6$ eV) play a significant
role in the chemistry and the heating. The first detection of a PDR in
an LBV nebula, through the presence of fine structure lines, was made
by Umana et al. (\cite{umana}) in their Spitzer study of the nebula that
surrounds HR Car. One year later a PDR was found in the nebula around the
LBV candidate HD 168625 (Umana et al. \cite{umana2}), this time through
spectral features indicating the presence of polycyclic aromatic
hydrocarbons (PAHs).

The three infrared fine structure lines mentioned above can be used to
determine the physical conditions in the PDR. But before that, any
possible contribution of the \ion{H}{ii} region to the observed line
intensities must be determined and subtracted. Neutral oxygen can be
found only in neutral regions, because its ionization potential (13.62 eV)
is very close to the ionization potential of hydrogen. Consequently,
the lines $[\ion{O}{i}]$ 63, 146\ \mbox{$\mu$m} arise exclusively from
the PDR (Malhorta et al. \cite{mal01}). However, carbon is the fourth-most
abundant element and has an ionization potential (11.26 eV) lower than
that of hydrogen, so that $\mathrm{C}^{+}$ can be found both in PDRs
and \ion{H}{ii} regions. Therefore, the line $[\ion{C}{ii}]$ 158\
\mbox{$\mu$m} may arise from the \ion{H}{ii} region of the nebula WRAY
15-751 and/or from the associated PDR (Heiles \cite{heil94}).

A first estimate of the contribution of the PDR to the flux of the
line $[\ion{C}{ii}]$ 158\ \mbox{$\mu$m} can be obtained following the
empirical method described by Goicoechea et al. (\cite{goi04}).  For
each spaxel where the $[\ion{N}{ii}]$ 122\ \mbox{$\mu$m} is detected
(see Appendix A), the $[\ion{C}{ii}]$ 158\ \mbox{$\mu$m} emission
that comes from the ionized gas should scale with the
$[\ion{N}{ii}]$ 122\ \mbox{$\mu$m}, since the latter arises
exclusively in ionized regions.  Fig.~\ref{pdr_correlation} shows the
correlation between the $[\ion{C}{ii}]$ 158\ \mbox{$\mu$m} and the
$[\ion{N}{ii}]$ 122\ \mbox{$\mu$m} flux for each spaxel where these
two lines are detected. This correlation is described by
      \begin{equation}
         F_{[\ion{C}{ii}]158} = (0.16\pm 0.02) \, F_{[\ion{N}{ii}]122}+ (0.03 \pm 0.01),
      \end{equation}
where $F_{[\ion{C}{ii}]158}$ is the $158\ \mbox{$\mu$m}$ line flux and
$F_{[\ion{C}{ii}]122}$ is the $122\ \mbox{$\mu$m}$ line flux in units of
10$^{-15}$ W~m$^{-2}$.  The constant term of this relation represents
the average $[\ion{C}{ii}]$ 158\ \mbox{$\mu$m} flux per
spaxel that arises in the PDR. Assuming that the PDR extends as the
dust nebula, i.e., over 18\arcsec\ in radius or $\sim$ 11 spaxels, we
then find $F_{[\ion{C}{ii}]158}^{\mathrm{PDR}} \simeq (0.33 \pm 0.11)\times
10^{-15}$W~m$^{-2}$.

Another estimate of the contribution of the PDR and
the \ion{H}{ii} regions to the flux of $[\ion{C}{ii}]$ 158\
\mbox{$\mu$m} line can be also obtained.
As the $[\ion{N}{ii}]$ 122\ \mbox{$\mu$m} line arises exclusively in
the ionized gas regions, measurements of its flux can give an estimate
of the contribution of the \ion{H}{ii} region to the flux of
$[\ion{C}{ii}]$ 158\ \mbox{$\mu$m} line,
$F_{[\ion{C}{ii}]158}^{\ion{H}{ii}}$ through a model. We define
$F_{[\ion{C}{ii}]158}^{\ion{H}{ii}} =\alpha F_{[\ion{C}{ii}]158}$,
where $F_{[\ion{C}{ii}]158}$ is the total flux of the $[\ion{C}{ii}]$
158\ \mbox{$\mu$m} line from Table~\ref{table:2} and $\alpha$ a factor
that has to be determined.
The ratio of fractional ionization is given, as previously, by 
      \begin{equation}
          \frac{\langle \mathrm{C}^{+}\rangle}{\langle \mathrm{N}^{+}\rangle}=
                 \frac{F_{[\ion{C}{ii}]158}^{\ion{H}{ii}}/\varepsilon_{[\ion{C}{ii}]158}}
                      {F_{[\ion{N}{ii}]122}/\varepsilon_{[\ion{N}{ii}]122}} \; .
      \end{equation}
Malhorta et al. (\cite{mal01}) provided an estimate for this relation only
for the high, $n_\mathrm{e}\gg n_\mathrm{crit}$, and the low,
$n_\mathrm{e}\ll n_\mathrm{crit}$, electron density limit, where
$n_\mathrm{crit}=3.1 \times 10^2\mathrm{cm}^{-3}$ for $[\ion{N}{ii}]$
122\ \mbox{$\mu$m} and $n_\mathrm{crit}=50 \mathrm{cm}^{-3}$ for
$[\ion{C}{ii}]$ 158\ \mbox{$\mu$m}. As neither of these two limits
apply to our case, the emissivities were calculated using the package
\textit{nebular} for the assumed $T_\mathrm{e}$ and the measured
$n_\mathrm{e}$. Assuming $\langle
\mathrm{C}^{+}\rangle/\langle \mathrm{N}^{+}\rangle=$ C/N, we find 
      \begin{equation} \label{eq:cnratio}
          \frac{F_{[\ion{C}{ii}]158}^{\ion{H}{ii}}}{F_{[\ion{N}{ii}]122}}=
                              (0.45 \pm 0.06) \ \frac{\mathrm{C}}{\mathrm{N}} \; .
      \end{equation} 
Since N/O has been estimated to be 1, we find that
      \begin{equation} \label{eq:cnratio2}
          \log\alpha = \log\frac{\mathrm{C}}{\mathrm{O}} + 0.31 \; ,
      \end{equation}
using the observed ratio $F_{[\ion{C}{ii}]158}/F_{[\ion{N}{ii}]122}$
= 0.222 $\pm$ 0.054.

\begin{figure*}[t]
\centering
   \includegraphics[width=16cm]{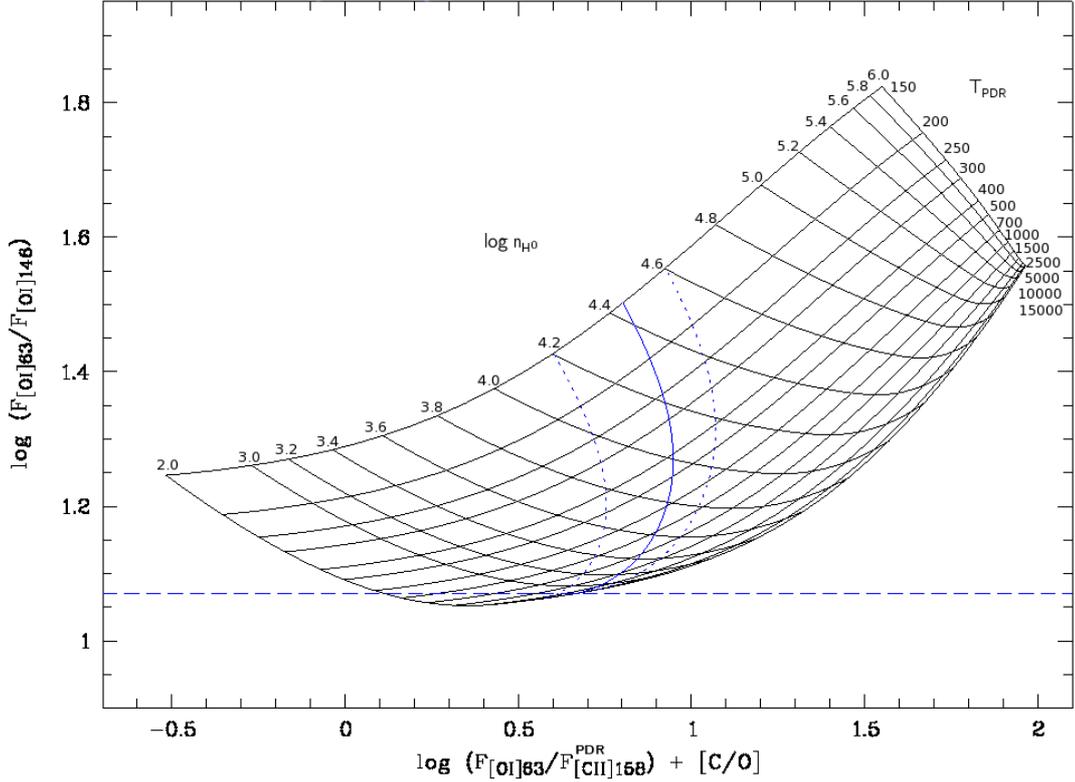}
     \caption{Temperature-density PDR diagnostic diagram. The grid
of flux ratios $F_{[\ion{O}{i}]63}/ F_{[\ion{O}{i}]146}$ versus
$F_{[\ion{O} {i}]63}/F_{[\ion{C}{ii}]158}^{\mathrm{PDR}}$ was
calculated by solving the level population equations for a range of
temperatures and densities. $F_{[\ion{O} {i}]63}/F_{[\ion{C}{ii}]158}
^{\mathrm{PDR}}$ is normalized to the solar abundance (C/O)$_{\odot}$ = 0.5
so that [C/O] $\equiv$ log(C/O) - log(C/O)$_{\odot}$. The solid line corresponds
to the pressure equilibrium between the \ion{H}{ii} region and the PDR, the two
dotted lines on each side accounting for the errors. The
horizontal dotted line represents the 1-$\sigma$ upper limit of the observational
log$(F_{[\ion{O}{i}]63}/F_{[\ion{O}{i}]146})$ ratio.}
     \label{lineratio}
\end{figure*}

To derive the temperature and density of the PDR as well as
the C/O abundance ratio, we plot the theoretical $F_{[\ion{O}{i}]63}/
F_{[\ion{O}{i}]146}$ ratio against the $F_{[\ion{O}
{i}]63}/F_{[\ion{C}{ii}]158}^{\mathrm{PDR}}$ ratio normalized to the
solar (C/O)$_{\odot}$ = 0.5 abundance ratio (Fig.~\ref{lineratio}),
following a similar study by Liu et al. (\cite{liu01}). To calculate
the populations of the fine-structure levels of C$^{+}$ and O$^{0}$,
we solved the two- and three-level atom equilibrium equations,
respectively, considering that collisions with atomic hydrogen
dominate in the PDR (Draine \cite{draine11}). The radiative transition
probabilities, A$_{ij}$, for the $[\ion{C}{ii}]$ and $[\ion{O}{i}]$
fine-structure lines and the electron collision strengths,
$\Omega_{ij}$, were taken from Draine (\cite{draine11}). The
collisional rate coefficients for the fine-structure excitation by
hydrogen were taken from Barinovs et al. (\cite{bar05}) for
$[\ion{C}{ii}]$ and from Abrahamsson et al. (\cite{abrah07}) for
$[\ion{O}{i}]$. A simple analytic extrapolation was made for
temperatures higher than those given in these two
references\footnote{The difference between the diagram of
Fig.~\ref{lineratio} and the one in the study of Liu et
al. (\cite{liu01}) is due to the use of updated collision
coefficients.}.  Furthermore, if we assume that there is pressure
equilibrium between the \ion{H}{ii} region and the
PDR, we have (Tielens \cite{tie05})
      \begin{equation} \label{eq:equilibrium}
        n_\mathrm{H^0}kT_\mathrm{PDR}\simeq 2n_\mathrm{e}kT_\mathrm{e}
              =(4.2\pm1.8)\times10^6\ \mathrm{cm}^{-3}\mathrm{K},
      \end{equation}
where $n_\mathrm{H^0}$ is the atomic hydrogen number density and
$T_\mathrm{PDR}$ is the temperature of the PDR. This relation defines
a locus of possible values in the diagram of Fig.~\ref{lineratio}.

Given the observed ratio $F_{[\ion{O}{i}]63}/F_{[\ion{O}{i}]146}$ =
8.4$\pm$2.8 and the constraints from Eq.~\ref{eq:equilibrium}, we
can derive from Fig.~\ref{lineratio} log$(F_{[\ion{O} {i}]63}/F_
{[\ion{C}{ii}]158}^{\mathrm{PDR}})$ + [C/O] = 0.63, where [C/O]
$\equiv$ log(C/O) - log(C/O)$_{\odot}$. Recalling that
$F_{[\ion{C}{ii}]158}^{\mathrm{PDR}}=(1-\alpha)F_{[\ion{C}{ii}]158}$
and considering the above relation between C/O and $\alpha$
(Eq.~\ref{eq:cnratio2}), the observed value of the line ratios yields
$\alpha$ = 0.82 $\pm$ 0.07 and C/O = 0.40$\pm$ 0.19. Considering the
errors, the C/O abundance ratio of the nebula has the solar value.
C/H is then $(1.7 \pm 1.3) \times 10^{-4}$ from the N/H, C/O and N/O
abundance ratios. The contribution of the \ion{H}{ii} region to
$[\ion{C}{ii}]$ 158\ \mbox{$\mu$m} is then $F_{[\ion{C}{ii}]158}^
{\ion{H}{ii}} =(0.99\pm0.16)\times 10^{-15}$W~m$^{-2}$ while the
contribution of the PDR is $F_{[\ion{C}{ii}]158}^{\mathrm{PDR}} = 
(0.22\pm0.09)\times 10^{-15}$W~m$^{-2}$. This value agrees with
the one obtained using the empirical method.

The diagram in Fig.~\ref{lineratio} also provides us with the values
of the density and the temperature of the PDR of the nebula, from the
observed $F_{[\ion{O}{i}]63}/F_{[\ion{O}{i}]146}$ ratio: $\log n_
\mathrm{H^0}$ = 2.38 $\pm$ 0.18 and $T_\mathrm{PDR} > 4000$ K. Given
the constraints from Eq.~\ref{eq:equilibrium}, we estimate that
$T_\mathrm{PDR} \sim 17500$ K but this value is very uncertain,
within a factor of 2.

The incident FUV radiation field, $G_0$, along with the density
$n_\mathrm{H^0}$, describes the structure of the PDR. Expressed in terms
of the average interstellar radiation field, which corresponds to a
unidirectional radiation field of $1.6 \times 10^{-3}$ erg cm$^{-2}$
s$^{-1}$, it is given by 
(Tielens \cite{tie05}) 
      \begin{equation}
          G_0 = 625\frac {L_{\star}\chi}{4\pi R^2}
      \end{equation}
at the distance $\textit{R}$ from the star, with $L_{\star}$ the
stellar luminosity and $\chi$ the fraction of the luminosity above 6
eV. For an early B star, $\chi\sim0.7$ (Young Owl et al.
\cite{young02}). Considering $L_{\star}=10^{5.8}L_{\odot}$ (Sect. 5.1)
and the radius of the ionized gas region, which is surrounded by the
PDR, $\textit{R}$ = 0.32 pc, the incident FUV radiation field is found
to be $G_0\simeq8.5\times10^4$ for the PDR of the WRAY 15-751
nebula. This value is consistent with the estimated PDR density, as
our results are reasonably compatible with the diagnostic diagrams of
the PDR models of Kaufman et al. (\cite{kaufman2}, Figs. 4 \& 5).

The FUV radiation given by $G_0$ is also absorbed and re-emitted by
the dust in the FIR.  The radiative equilibrium gives us the dust
temperature, $T_\mathrm{dust}$, which in case of silicates
(i.e., $\beta=2$) is given by (Tielens \cite{tie05})
      \begin{equation}
          T_\mathrm{dust}=50\left(\frac{1\mu\mathrm{m}}{a}\right)^{0.06}
                          \left(\frac{G_0}{10^4}\right)^{1/6} \mathrm{K\ \
                         for}\ T_\mathrm{dust} < 250\ \mathrm{K} \; .
      \end{equation}
As the small grains dominate the average cross-section, a typical
grain size of $a = 0.1\ \mu \mathrm{m}$ can be assumed, which
leads to a dust temperature of $T_\mathrm{dust}$=81 K, in excellent
agreement with the results of the 2-Dust model.

The total mass of hydrogen in the PDR, $M_{\mathrm{H}}$, can be
estimated from the $[\ion{C}{ii}]$ 158\ \mbox{$\mu$m} line
flux derived for the PDR (Tielens \cite{tie05}), using the equation
given in Appendix C. For the above PDR density, temperature,
distance and C/H abundance, the neutral hydrogen mass in the PDR is
estimated to be $M_{\mathrm{H}}= 0.43 \pm 0.35\ \mathrm{M}_{\odot}$.

\subsection{Total gas mass}

The total gas mass of the nebula is the sum of the mass in the 
ionized nebula and the mass in the PDR, corrected for the presence of helium, 
i.e.,
      \begin{equation}
          M_{\mathrm{gas}}=(1+4y)\,(M_{i}+
                   M_{\mathrm{H}}),
      \end{equation}
where $y=n_{\mathrm{He}}/n_{\mathrm{H}}$. Assuming a solar abundance
for helium of 12+log(He/H)=10.93$\pm$0.01 (Grevesse et
al. \cite{grevesse}), the gas mass is $M_{\mathrm{gas}}$=1.7$\pm$0.6
M$_{\odot}$. Considering the calculated dust mass (Sect. 5.1), the
dust-to-gas mass ratio for the inner nebula is
$M_{\mathrm{dust}}/M_{\mathrm{gas}}$=0.026$\pm$0.010 , i.e., $\sim$
3\%.  If the He abundance is higher, as expected for an evolved star,
the total gas mass will be higher, typically 20\% for a He/H abundance
ratio corresponding to the observed N/O abundance ratio.

\section{Discussion}

\begin{table}[t]
\caption{Parameters of WRAY 15-751}
\label{table:4}
\begin{tabular}{llc}\hline\hline\\[-0.10in]                          
Star & log $L/L_{\odot}$                       &  5.7 $\pm$ 0.2                   \\
     & $T_{\mathrm{eff}}$ (K)                    &  30000 $\leftrightarrow$ 9000   \\
     & $D$ (kpc)                             &  6.0 $\pm$ 1.0  \\
Inner Shell & \textit{r}  (pc)               &  0.5                             \\
     & ${\rm v}_{\mathrm{exp}}$ (km s$^{-1}$)     & 26                               \\
     & $t_{\mathrm{kin}}$  (10$^{4}$ yr)          & 1.9                              \\
     & $n_{\mathrm{e}}$ (cm$^{-3}$)               &  210 $\pm$ 80                    \\
     & $T_{\mathrm{e}}$  (K)                     &  10000                          \\
     & N/O                                    &  1.00 $\pm$ 0.38                   \\
     & C/O                                    &  0.40 $\pm$ 0.19                 \\
     & 12+log N/H                             &  8.63 $\pm$ 0.20                 \\
     & $M_{\mathrm{dust}}$  (10$^{-2}$ M$_{\odot}$)   &  4.5 $\pm$ 0.5         \\
     & $M_{\mathrm{gas}}$   (M$_{\odot}$)           &  1.7 $\pm$ 0.6         \\
Outer Shell & \textit{r}  (pc)                 &  2.0                             \\
     & $t_{\mathrm{kin}}$ (10$^{4}$ yr)            & 7.5                              \\
     & $M_{\mathrm{dust}}$ (10$^{-2}$ M$_{\odot}$)   &  5.0 $\pm$ 2.0         \\
\hline
\end{tabular}
\end{table}

A summary of the measurements obtained in the previous sections is
given in Table~\ref{table:4}.  The luminosity, effective temperature
and distance of the central star are given first (from Hu et
al.\cite{hu}, Sterken and al. \cite{ste08}, and this work),
followed by the parameters of the inner and the outer dust shells, i.e.,
the radius, the expansion velocity (from HVD; assumed to be identical
for both shells), the kinematic age, the electron density and the assumed
temperature of the ionized gas, the abundance ratios, and the dust and gas
masses.

The Herschel-PACS far-infrared images of WRAY 15-751 reveal the dust
nebula as a shell of radius 0.5 pc and width 0.35 pc.  These
observations also unveiled a second dust nebula, four times bigger,
lying in an empty cavity. As in the case of WR stars (Marston
\cite{mar96}), the empty cavity probably corresponds to the interior of
O-star wind bubble formed when the star was on the main sequence.

Our study consistently shows that the main nebula is illuminated by an
average early-B star and consists of a shell of ionized gas
surrounded by a thin photodissociation region. Both these regions are
mixed with dust. The mass of this nebula amounts to $\sim$ 2
M$_{\odot}$ ejected $\sim$ 2 $\times$ 10$^4$ years ago. The second,
larger and older nebula contains a similar amount of mass if we
assume a similar dust-to-gas ratio so that, in total, $\sim$ 4
M$_{\odot}$ of gas have been ejected within $\sim$ 6 $\times$ 10$^4$
years. This also indicates that the star had multiple episodes of
intense mass-loss. Moreover, it is possible that the very inner
dense nebula spectroscopically detected by HVD and resolved by Duncan
and White (\cite{duncan}) constitutes a third ejection some 10$^3$
years ago, assuming an angular radius of 1$\arcsec$ (Duncan and White
\cite{duncan}) and an expansion velocity of 26 km s$^{-1}$, the same
as for the main ring nebula.

The N/O abundance ratio appears to be enhanced by a factor 8 with respect to
the solar abundances given in Ekstrom et al. (\cite{eks12}). This
confirms the presence of processed material in the nebula. The C/O
ratio, measured for the first time in a LBV nebula, is solar within
the uncertainties.  These ratios correspond to an enhancement in N/H
by a factor 6 and a depletion in C/H and O/H by a factor 1.4 with
respect to the solar abundances.

The N/O ratio of 1.00 $\pm$ 0.38 is quite similar to the ratio measured
in the nebula around the LMC LBV R127 (Smith et al. \cite{smith98}).
The 12+log(N/H) abundance of 8.63 $\pm$ 0.20 is between the values for
the LBVs AG Car and $\eta$ Car (Smith et al. \cite{smith98}).  The
conclusion of Smith et al. (\cite{smith97_2}, \cite{smith98}) that LBV
nebulae contain only mildly enriched material with respect to CNO
equilibrium values and thus were ejected during a previous red
supergiant (RSG) or yellow supergiant (YSG) phase therefore applies to
WRAY 15-751, especially as the star is less luminous, log
$L/L_{\odot}$ = 5.7 $\pm$ 0.2, i.e., just at or below the
Humphreys-Davidson limit. In addition, the nebula has a relatively low
expansion velocity of 26 km/s, more compatible with RSG outflows than
the higher velocities measured in more luminous LBVs such as AG Car.  The
ejection of the WRAY 15-751 nebula during an RSG phase was also
proposed by Voors et al. (\cite{voo00}) on the basis on its dust
composition.

\begin{figure}[t]
\resizebox{\hsize}{!}{\includegraphics*{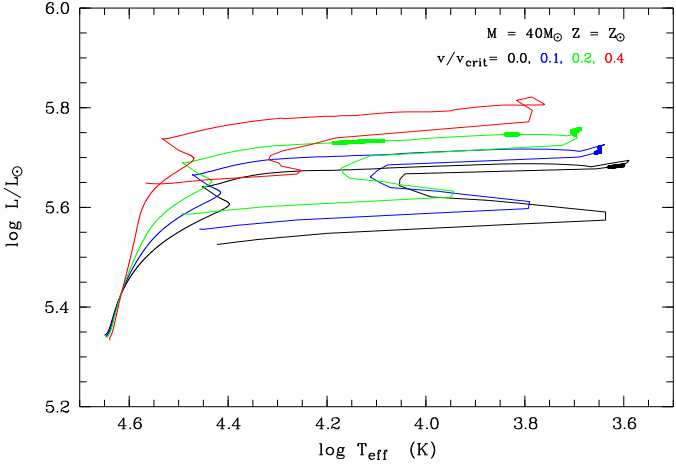}}
\caption{Evolutionary path in the HR diagram of a 40 M$_{\odot}$
star of solar metallicity and for initial rotation rates
$\rm v/\rm v_{crit}$ from 0 to 0.4, using the models of Ekstr\"om et
al. (\cite{eks12}). The thicker lines emphasize the part of the
tracks compatible with the measurements (cf. Fig.~\ref{nso_cso} and
Fig.~\ref{nso_mdot}). For clarity, the tracks are
stopped at the beginning of the blue loop (data point n$^{\rm o}$ 210 in
Ekstr\"om et al. \cite{eks12}).}
\label{hr}
\end{figure}

\begin{figure}[t]
\resizebox{\hsize}{!}{\includegraphics*{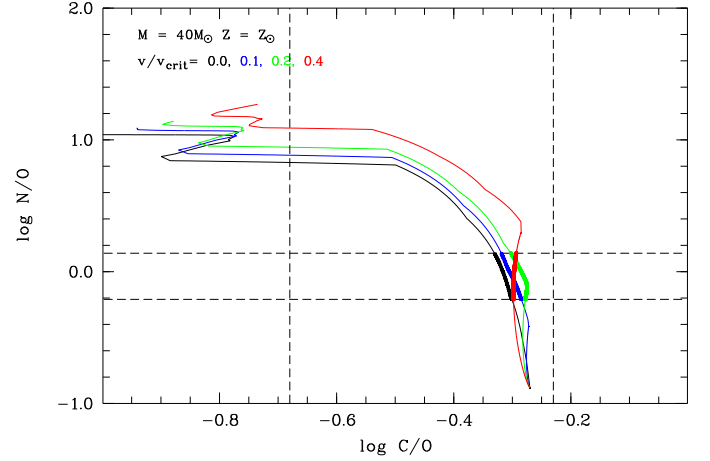}}
\caption{Evolution of the N/O versus the C/O surface abundance
ratios for a 40 M$_{\odot}$ star of solar metallicity and for initial
rotation rates $\rm v/\rm v_{crit}$ from 0 to 0.4, using the models
of Ekstr\"om et al. (\cite{eks12}). The dashed lines correspond to the
values measured for the inner shell around WRAY 15-751, with their
errors.  The thicker lines emphasize the part of the tracks compatible
with the measurements. For clarity, the tracks are stopped at the
beginning of the blue loop (data point n$^{\rm o}$ 210 in Ekstr\"om et
al. \cite{eks12}).}
\label{nso_cso}
\end{figure}

\begin{figure}[t]
\resizebox{\hsize}{!}{\includegraphics*{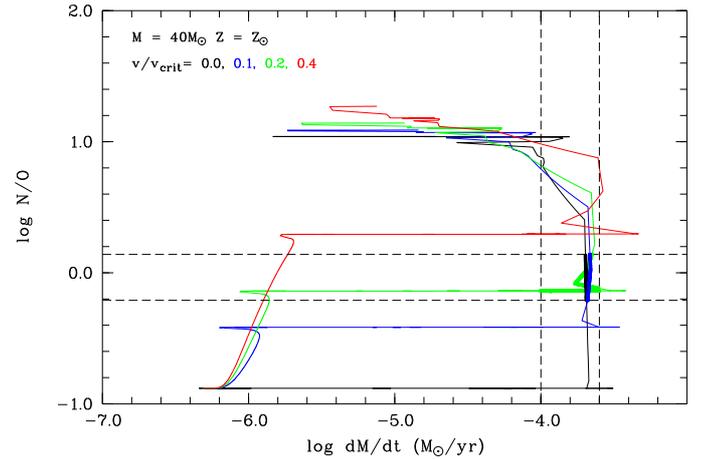}}
\caption{Evolution of the N/O surface abundance ratio as a
function of the mass-loss rate for a 40 M$_{\odot}$ star of solar
metallicity and for initial rotation rates $\rm v/\rm v_{crit}$ from
0 to 0.4, using the models of Ekstr\"om et al. (\cite{eks12}).  The
dashed lines correspond to the values measured for the inner shell
around WRAY 15-751, with their errors.  The thicker lines emphasize the
part of the tracks compatible with the measurements. For clarity, the
tracks are stopped at the beginning of the blue loop (data point
n$^{\rm o}$ 210 in Ekstr\"om et al. \cite{eks12}).
}
\label{nso_mdot}
\end{figure}

Our observations can be used to constrain the evolutionary path of the
star and the epoch of ejection of the nebula.  Given its luminosity,
WRAY 15-751 is expected to result from the evolution of a star of
initial mass in the range 40 -- 60 M$_{\odot}$.  Figs.~\ref{hr},
\ref{nso_cso}, and \ref{nso_mdot} show the evolution of some properties
of a 40 M$_{\odot}$ star using the models of Ekstrom et
al. (\cite{eks12}).  Four different cases of stellar rotation are
considered, from no rotation to a rotation rate of
$\rm v/\rm v_{crit}$=0.4. The tracks are first compared with the
observed N/O and C/O abundance ratio to identify the part of the
tracks where the computed surface abundances match the observed
nebular abundances (Fig.~\ref{nso_cso}). Possible tracks were
additionally constrained when compared with the observed mass-loss rate
(Fig.~\ref{nso_mdot}). The mass-loss rate is estimated from the mass
of the inner nebula divided by the duration of the enhanced mass-loss
episode (estimated from the kinematic age), i.e., the time needed to cover
the ring width: $\log \dot{M}$ = $-$3.8 $\pm$ 0.2,
accounting for an increase of the mass of 20\% due to the higher He/H
abundance at that time. These constraints are finally reported in the
HR diagram (Fig.~\ref{hr}) to identify the locus of the ejection of
the inner nebula. The model appears to remarkably agree with the
observations, supporting the scenario of an ejection of the nebula during
the RSG phase\footnote{We loosely use the term RSG for the reddest part of
the tracks in the HR diagram, while strictly speaking RSG have $\log
T_{\rm eff} <$ 3.65.}.  The N/H enhancement factor as well as the C/H
and O/H depletion factors predicted by the model also agree with the
observed values. The mild N/O enrichment indicates that the star
cannot be a fast rotator, the $\rm v/\rm v_{crit}$=0.4 track being
clearly excluded. A similar result is obtained for a 45 M$_{\odot}$
star except that the only possible tracks have $\rm v/\rm v_{crit}$=
0.0 and 0.1. For a 50 M$_{\odot}$ star, no track satisfies the
observational constraints.

The time between the ejections of the outer and the inner nebulae,
about 6$\times$10$^4$ years, is compatible with the duration of the
RSG phase of a 40 M$_{\odot}$ star, as computed by the model.  Thus,
the outer nebula could also have been ejected during this phase,
especially in the models with $\rm v/\rm v_{crit} <$ 0.2 for which
the inner, younger nebula is ejected closer to the end of the RSG
phase than to the beginning (Fig.~\ref{hr}).  The total mass lost
during the RSG phase amounts to 8 - 9 M$_{\odot}$ in the
model. Although higher, this is compatible with our value of 4 $\pm$ 2
M$_{\odot}$, recalling that the value for the outer nebula is particularly
uncertain due to the unknown dust-to-gas ratio and He abundance.

Our results suggest that the ejection of the nebula does not occur
because the star is rotating close to the critical velocity, as
proposed by Meynet et al. (\cite{meyn11}). Moreover, the existence of
multiple nebular shells points to an instability mechanism at work
during the RSG evolutionary stage and not to a continuous wind.  In
particular, models by Stothers and Chin (\cite{stoth96}) suggest that
LBV nebulae can result from strong, closely spaced mass-loss episodes
in the RSG phase and not from a continuous wind. For a 45 M$_{\odot}$
star, they found that about 4 M$_{\odot}$ can be ejected, in agreement
with our measurements.

While our results support the scenario of an ejection of LBV in the RSG phase,
the study of Lamers et al. (\cite{lamers}) reached the conclusion that LBV
nebulae were ejected during the BSG phase with high rotational
velocities, and not during the RSG stage. However, Lamers et
al. (\cite{lamers}) only considered very luminous LBVs ($\log L / L
_{\odot} > 5.8$), while WRAY 15-751 is a lower luminosity
LBV. Moreover, the nebula around WRAY 15-751 is only weakly bipolar
compared with other LBV nebulae such as those around AG Car or HR Car, in
qualitative agreement with little effect of rotation. Finally, the
discovery of a dusty LBV-like ring nebula around the yellow supergiant
Hen3-1379, which is very similar to WRAY 15-751, also supports the ejection
of nebulae during the RSG phase (Hutsem\'ekers et al. \cite{hut13}).
Therefore, high-luminosity and low-luminosity LBVs probably follow
different evolutionary paths.


Our results are compatible with the evolutionary model of an $\sim$
40 M$_{\odot}$ star and the O--BSG--RSG--YSG--LBV filiation.
According to Toal\'a and Arthur (\cite{toa11}), an $\sim$ 40
M$_{\odot}$ star creates a bubble of radius $\sim$ 25 pc as a main-
sequence O star, in agreement with the structure tentatively observed
in Fig.~\ref{higal}. Then, when an RSG, the star ejects several solar
masses of material in the cavity previously created, forming the
observed dusty nebulae. It is interesting to note that in this
scenario, the age of WRAY 15-751 since the ejection of the last nebula
is only $\sim$ 2 $\times$ 10$^{4}$ years, which corresponds in the
computed tracks of Fig.~\ref{hr} to the loop at $\log T_{\rm eff} \sim $
4.1. Higher temperatures are only reached $\sim$ 10$^{5}$ years
later. It is not clear whether a star at that location in the HR
diagram, which corresponds to a hot YSG, can have the LBV-like
instability properties currently displayed by WRAY 15-751. This might
indicate that the LBV phenomenon could occur at different evolutionary
stages. Determining the surface abundances of WRAY 15-751 in its
present stage might help to constrain this scenario more closely.

\section{Conclusions}

We have presented the analysis of Hershel PACS imaging and
spectroscopic data of the nebula around the LBV Wray 15-751, together
with new optical-imaging data. The far-infrared images clearly show
that the main, dusty nebula is a shell extending outside the
well-known H$\alpha$ nebula.  Furthermore, these images reveal a
second, bigger and fainter dust nebula that is observed for the first
time. The two nebulae lie in an empty cavity, very likely the remnant of the
O-star wind bubble formed when the star was on the main sequence.

The dust parameters of the main nebula were determined based on dust
modeling.  This model shows that the far-infrared emission did not
significantly change during the different phases of the S Dor
cycle. This stability points to a stellar variation under essentially
constant luminosity. We also found that Fe-rich dust is needed to
reproduce the data. This is not unexpected in LBV nebulae as a
consequence of depletion of C and O with respect to heavier elements
(Gail et al. \cite{gail05}).

The far-infrared spectrum of the main nebula contains forbidden
emission lines coming from an ionized region and from a
photodissociation region, from which we derived the gas parameters,
such as the C, N, O abundances and the ejected gas mass, with the C/O
ratio measured for the first time in an LBV nebula. As a result
of this study, the main shell nebula consists of an ionized gas region
which is surrounded by a thin PDR, both regions being mixed with the
dust.  As expected for such an evolved star, the nebula shows N
enrichment and C, O depletion.

The measured abundances, masses and kinematic ages of the nebulae were
used to constrain the evolution of the star and the epoch at which the
nebulae were ejected.  Our results point to an ejection of the nebulae
during the RSG evolutionary phase of an $\sim$ 40 M$_{\odot}$ star.
The multiple shells around the star suggest that the mechanism of
mass-loss is not a continuous wind but instead a series of short episodes of
extreme mass-loss.

This scenario is compatible with the recent evolutionary tracks
computed for an $\sim$ 40 M$_{\odot}$ star with little rotation, in
particular the O--BSG--RSG--YSG--LBV filiation although it
should be stressed that post-main-sequence evolutionary tracks of
massive stars are still very uncertain, in particular since they rely
on poorly known mass-loss mechanisms and rates. If the evolutionary
tracks are correct, our results support the idea that high-luminosity and
low-luminosity LBVs follow different evolutionary paths. The forthcoming
analysis of similar data for higher luminosity LBVs (e.g., AG Car) and for
LBVs known to be fast rotators (e.g., HR Car) should allow us to
constrain this scenario more closely.

\begin{acknowledgements}
We thank Xiaowei Liu, Daniel Pequignot and Evelyne Roueff for help with
the PDR diagnostic diagram construction. CVN, PR, DH, YN, KE and MATG
acknowledge support from the Belgian Federal Science Policy Office via
the PRODEX Programme of ESA. The Li\`ege team acknowledges also
support from the FRS-FNRS (Comm. Fran{\c c}. de Belgique).  PACS has
been developed by a consortium of institutes led by MPE (Germany) and
including UVIE (Austria); KU Leuven, CSL, IMEC (Belgium); CEA, LAM
(France); MPIA (Germany); INAF-IFSI/OAA/OAP/OAT, LENS, SISSA (Italy);
IAC (Spain).  This development has been supported by the funding
agencies BMVIT (Austria), ESA-PRODEX (Belgium), CEA/CNES (France), DLR
(Germany), ASI/INAF (Italy), and CICYT/MCYT (Spain). Data presented in
this paper were analyzed using “HIPE”, a joint development by the
Herschel Science Ground Segment Consortium, consisting of ESA, the
NASA Herschel Science Center, and the HIFI, PACS and SPIRE
consortia. This research has made use of the NASA/IPAC Infrared
Science Archive, which is operated by the Jet Propulsion Laboratory,
California Institute of Technology.
\end{acknowledgements}

\clearpage

\begin{appendix}
  \section{Emission line fluxes for each spaxel}

\begin{table*}[h]
\caption{Line fluxes in each spaxel. A dash indicates a poor S/N or a non-detection. The spatial configuration corresponds to the footprint of the PACS-spectrometer as it is displayed in Fig.~\ref{wra_foot}.}
\label{table:5}
\centering
\begin{tabular}{c c| c c| c c| c c | c c}
\hline\hline                            \\
Ion               & $\lambda$ (band)  & $F (\times10^{-15})$ & $\Delta F (\times10^{-15})$ & $F (\times10^{-15})$ & $\Delta F (\times10^{-15})$ & $F (\times10^{-15})$ & $\Delta F (\times10^{-15})$  &$F (\times10^{-15})$ &  $\Delta F (\times10^{-15})$\\
                  & $(\mu m)$  & (W~m$^{-2}$)          & (W~m$^{-2}$)                 & (W~m$^{-2}$)          & (W~m$^{-2}$)                 & (W~m$^{-2}$)          & (W~m$^{-2}$)                  & (W~m$^{-2}$)  & (W~m$^{-2}$)\\
\hline\hline
                 &             & $\underline{spaxel\ 4,4}$   &           & $\underline{spaxel\ 4,3}$  &             & $\underline{spaxel\ 4,2}$  &         & $\underline{spaxel\ 4,1}$   &            \\
$[\ion{N}{iii}]$ & 57 (B2A)    &      -       &    -      &     -       &     -       &      -      &    -    &      -       &     -     \\
$[\ion{O}{i}]$   & 63 (B2A)    &      -       &    -      &     -       &     -       &     -       &    -    &      -       &     -      \\
$[\ion{N}{iii}]$ & 57 (B3A)    &      -       &     -     &     -       &     -       &      -      &    -    &       -      &     -      \\
$[\ion{O}{i}]$   & 63 (B3A)    &      -       &     -     &     -       &     -       &      -      &    -    &       -      &     -      \\
$[\ion{O}{iii}]$ & 88 (B2B)    &       -      &    -      &     -       &      -      & 0.13        & 0.03    &        -     &       -    \\
$[\ion{N}{ii}]$  & 122 (R1B)   &      -       &    -      &  0.09       & 0.02        & 0.10        & 0.02    &        -     &         -   \\
$[\ion{O}{i}]$   & 146 (R1B)   &      -       &    -      &     -       &      -      &     -       &     -   &     -        &            \\
$[\ion{C}{ii}]$  & 158 (R1B)   & 0.04         & 0.01      &  0.06       & 0.01        & 0.03        & 0.01    & 0.04         & 0.01       \\
$[\ion{C}{ii}]$  & 158 (R1A)   & 0.05         & 0.01      &  0.05       & 0.01        & 0.04        & 0.01    &     -        &      -      \\
$[\ion{N}{ii}]$  & 205 (R1A)   &      -       &   -       &    -        &      -      &    -        &    -    &     -        &      -     \\
\hline
                 &             & $\underline{spaxel\ 3,4}$   &          & $\underline{spaxel\ 3,3}$   &             & $\underline{spaxel\ 3,2}$  &         & $\underline{spaxel\ 3,1}$   &             \\
$[\ion{N}{iii}]$ & 57 (B2A)    &    -         &   -      & 0.29         & 0.08        & 0.24        & 0.09    &     -        &             \\
$[\ion{O}{i}]$   & 63 (B2A)    &    -         &   -      & 0.15         & 0.06        & 0.26        & 0.06    &     -        &      -      \\
$[\ion{N}{iii}]$ & 57 (B3A)    &    -         &   -      & 0.16         & 0.08        & 0.18        & 0.07    &     -        &      -      \\
$[\ion{O}{i}]$   & 63 (B3A)    &    -         &   -      & 0.11         & 0.07        & 0.13        & 0.06    &     -        &      -      \\
$[\ion{O}{iii}]$ & 88 (B2B)    &    -         &   -      & 0.10         & 0.04        & 0.10        & 0.03    & 0.10         & 0.04        \\
$[\ion{N}{ii}]$  & 122 (R1B)   & 0.06         & 0.02     & 0.50         & 0.02        & 0.71        & 0.02    & 0.04         & 0.02         \\
$[\ion{O}{i}]$   & 146 (R1B)   &    -         &   -      &    -         &    -        &     -       &    -    &     -        &      -       \\
$[\ion{C}{ii}]$  & 158 (R1B)   & 0.07         & 0.01     & 0.11         & 0.01        & 0.14        & 0.01    & 0.02         & 0.01         \\
$[\ion{C}{ii}]$  & 158 (R1A)   & 0.05         & 0.01     & 0.14         & 0.01        & 0.16        & 0.01    & 0.04         & 0.01         \\
$[\ion{N}{ii}]$  & 205 (R1A)   &      -       &   -      &    -         &      -      &      -      &    -    &     -        &     -        \\
\hline
                 &             & $\underline{spaxel\ 2,4}$  &          & $\underline{spaxel\ 2,3}$    &             & $\underline{spaxel\ 2,2}$  &         & $\underline{spaxel\ 2,1}$   &              \\
$[\ion{N}{iii}]$ & 57 (B2A)    &     -       &    -     & 0.18          & 0.08        & 0.25        & 0.08    & 0.23         & 0.08       \\
$[\ion{O}{i}]$   & 63 (B2A)    &     -       &    -     & 0.17          & 0.06        & 0.40        & 0.06    &    -         &   -          \\
$[\ion{N}{iii}]$ & 57 (B3A)    &     -       &    -     & 0.27          & 0.08        & 0.25        & 0.07    &    -         &   -          \\
$[\ion{O}{i}]$   & 63 (B3A)    &     -       &    -     & 0.14          & 0.06        & 0.35        & 0.06    &    -         &   -          \\
$[\ion{O}{iii}]$ & 88 (B2B)    &  0.12       & 0.04     & 0.08          & 0.04        & 0.14        & 0.04    & 0.11         & 0.04         \\
$[\ion{N}{ii}]$  & 122 (R1B)   &  0.11       & 0.02     & 0.74          & 0.02        & 0.71        & 0.02    & 0.27         & 0.02         \\
$[\ion{O}{i}]$   & 146 (R1B)   &     -       &    -     &     -         &      -      & 0.02        & 0.01    &    -         &    -         \\
$[\ion{C}{ii}]$  & 158 (R1B)   &  0.06       & 0.01     & 0.15          & 0.01        & 0.15        & 0.01    & 0.06         & 0.01         \\
$[\ion{C}{ii}]$  & 158 (R1A)   &  0.07       & 0.01     & 0.16          & 0.01        & 0.16        & 0.01    & 0.07         & 0.01         \\
$[\ion{N}{ii}]$  & 205 (R1A)   &      -      &     -    & 0.17          & 0.03        & 0.26        & 0.04    &     -        &     -         \\
\hline
                 &             & $\underline{spaxel\ 1,4}$  &          & $\underline{spaxel\ 1,3}$    &             & $\underline{spaxel\ 1,2}$  &        & $\underline{spaxel\ 1,1}$   &              \\
$[\ion{N}{iii}]$ & 57 (B2A)    &     -       &     -    & 0.12          & 0.08        & 0.23         & 0.08   & 0.17         & 0.08         \\
$[\ion{O}{i}]$   & 63 (B2A)    &     -       &     -    & 0.07          & 0.06        &      -       &    -   &     -        &    -          \\
$[\ion{N}{iii}]$ & 57 (B3A)    &     -       &     -    &      -        &    -        & 0.19         & 0.07   &     -        &    -          \\
$[\ion{O}{i}]$   & 63 (B3A)    &     -       &     -    &      -        &    -        & 0.18         & 0.06   &     -        &    -          \\
$[\ion{O}{iii}]$ & 88 (B2B)    &  0.08       & 0.04     &      -        &    -        & 0.11         & 0.03   & 0.07         & 0.02          \\
$[\ion{N}{ii}]$  & 122 (R1B)   &  0.06       & 0.02     & 0.45          & 0.02        & 0.51         & 0.02   & 0.24         & 0.02          \\
$[\ion{O}{i}]$   & 146 (R1B)   &     -       &     -    &      -        &     -       &     -        &   -    &     -        &    -          \\
$[\ion{C}{ii}]$  & 158 (R1B)   &  0.03       & 0.01     & 0.07          & 0.01        & 0.09         & 0.01   & 0.03         & 0.01           \\
$[\ion{C}{ii}]$  & 158 (R1A)   &  0.04       & 0.01     & 0.09          & 0.01        & 0.09         & 0.01   & 0.05         & 0.01           \\
$[\ion{N}{ii}]$  & 205 (R1A)   &      -      &    -     &    -          &    -        &     -        &   -    &     -        &     -          \\
\hline
                 &             & $\underline{spaxel\ 0,4}$  &          & $\underline{spaxel\ 0,3}$    &             & $\underline{spaxel\ 0,2}$   &        & $\underline{spaxel\ 0,1}$   &               \\
$[\ion{N}{iii}]$ & 57 (B2A)    &      -      &    -     & 0.17          & 0.08        &      -       &    -   &     -        &    -            \\
$[\ion{O}{i}]$   & 63 (B2A)    &      -      &    -     &      -        &     -       &      -       &    -   &     -        &    -            \\
$[\ion{N}{iii}]$ & 57 (B3A)    &      -      &    -     &      -        &     -       &      -       &    -   &     -        &    -            \\
$[\ion{O}{i}]$   & 63 (B3A)    &      -      &    -     &      -        &     -       &      -       &    -   &     -        &    -            \\
$[\ion{O}{iii}]$ & 88 (B2B)    &      -      &    -     &      -        &     -       &      -       &    -   &     -        &    -            \\
$[\ion{N}{ii}]$  & 122 (R1B)   &      -      &    -     & 0.11          & 0.02        & 0.13         & 0.02   &     -        &    -            \\
$[\ion{O}{i}]$   & 146 (R1B)   &      -      &    -     &      -        &      -      &      -       &     -  &     -        &    -             \\
$[\ion{C}{ii}]$  & 158 (R1B)   &      -      &    -     & 0.02          & 0.01        &      -       &     -  &     -        &    -             \\
$[\ion{C}{ii}]$  & 158 (R1A)   &      -      &    -     & 0.04          & 0.01        &      -       &     -  &     -        &    -             \\
$[\ion{N}{ii}]$  & 205 (R1A)   &     -       &    -     &     -         &      -      &     -        &     -  &     -        &     -            \\
\hline
\end{tabular}
\end{table*}

Table~\ref{table:5} gives the results of the emission line flux
measurements for each spaxel. The first column contains the detected
ions along with the spectral band in which the corresponding line was
measured. The following columns contain the line fluxes, expressed in
W/m$^{2}$, along with their errors. The spaxel numbers
(Fig.~\ref{wra_foot}) are mentioned in every cell of the table. No
spectral lines were detected in the western column of the
spectrometric camera (spaxels [0,0] to [4,0]), and these spaxels were
hence not included in the table. The quoted uncertainties are the sum
of the line-fitting uncertainty plus the uncertainty due to the
position of the continuum.

\section{Ionized nebula}
The formulae needed to estimate the nebular mass and the ionizing flux
from both H$\alpha$ and radio emissions are re-derived here for
consistency of hypotheses and notations.

\subsection{$\mathrm{H}\alpha$ emission}

The luminosity in the H$\alpha$ recombination line, integrated over the volume \textit{V} of the nebula, is
given by (Osterbrock \& Ferland, \cite{oster06})
      \begin{equation}
           L(\mathrm{H\alpha})=\int_V 4 \pi j_{\mathrm{H\alpha}}\epsilon \mathrm{d}V,
      \end{equation}
where $\epsilon$ is the filling factor that gives the fraction of the volume of the nebula that is filled
by ionized gas, and $j_{\mathrm{H\alpha}}$ is the H$\alpha$ line emission coefficient.
The flux received by the observer is 
      \begin{equation}
           F_{0}(\mathrm{H\alpha})=\frac{L(\mathrm{H\alpha})}{4\pi D^{2}},
      \end{equation}
where $\textit{D}$ is the distance to the nebula. By integrating over the volume, assuming a spherical uniform nebula of
radius $\textit{R}$ and considering the effective recombination coefficient 
$\alpha^{\mathrm{eff}}=(4\pi j)/(n_{\mathrm{e}}n_{\mathrm{p}}h\nu)$,
we have
      \begin{equation} \label{flux H alpha}
           F_{0}(\mathrm{H\alpha})=\left(\frac{R^{3}}{3 D^{2}}\right)\epsilon h \nu_{\mathrm{H}_{a}}n_{
                         \mathrm{e}}n_{\mathrm{p}} \alpha^{\mathrm{eff}}_{\mathrm{H}_{a}},
      \end{equation}
where $n_{\mathrm{e}}$ is the electron density, $n_{\mathrm{p}}$ is the proton density, $\textit{h}$ is the
Plank's constant and $\nu_{\mathrm{H}_{a}}$ is the frequency of the H$\alpha$ line.

The mass of the ionized nebula, $M_{i}$, is equal to
       \begin{equation} \label{ionized mass} 
           M_{i}=\frac{4\pi}{3}R^{3}\mu_{+} n_{\mathrm{p}}m_{\mathrm{H}}\epsilon,
      \end{equation}
with $m_\mathrm{H}$ being the H atomic mass and $\mu_{+}$ the mean ionic mass per H ion. By replacing 
$n_{\mathrm{e}}$ in the equation (\ref{flux H alpha}) with $n_{\mathrm{e}}=x_{e} n_{\mathrm{p}}$ and combining it
with equation (\ref{ionized mass}), the ionized mass can be written as
       \begin{equation} 
           M_{i(\mathrm{H\alpha})}=\frac{4\pi \mu_{+} m_{\mathrm{H}}} {\sqrt{3h\nu_{\mathrm{H}_{a}}x_{e}\alpha^{\mathrm{eff}}_
                     {\mathrm{H}_{a}}}}\epsilon^{1/2}\theta^{3/2}D^{5/2}F^{1/2}_0(\mathrm{H}\alpha),
      \end{equation}
where $\theta$ is the angular radius of the nebula ($R=\theta D$) in $\mathrm{H}\alpha$.
By replacing the effective recombination  coefficient with the following formula, taken from Draine
(\cite{draine11}),
      \begin{equation}
           \alpha^{\mathrm{eff}}_{\mathrm{H}_{a}}=1.17\times10^{-13}T_{4}^{(-0.942-0.031\mathrm{ln}T_{4})}\ 
                             \mathrm{cm^{3}s^{-1}},
      \end{equation}
where $T_{4}=T_\mathrm{e}/(10^{4}\ \mathrm{K})$ and $T_\mathrm{e}$ is the electron temperature, 
we obtain the following expression for the ionized mass
of the nebula in solar masses
       \begin{equation} \label{ionized mass H alpha}
              M_{i(\mathrm{H\alpha})}=57.9\frac{1+4y_{+}}{\sqrt{1+y_{+}}}T_4^{(0.471+0.015\mathrm{ln}T_4)}\epsilon^{1/2}
                   \theta^{3/2}D^{5/2}F^{1/2}_0(\mathrm{H}\alpha),
       \end{equation}
where $\theta$ is in arcsec, $\textit{D}$ is in kpc and
$F_0(\mathrm{H}\alpha)$ is in ergs~cm$^{-2}$~s$^{-1}$. With
$n_{\mathrm{H}^{+}}=n_{\mathrm{p}}$, $n_{\mathrm{He}^{+}}$ and
$n_{\mathrm{He}^{++}}$ the ionized hydrogen, ionized helium and doubly
ionized helium number densities, respectively,
$x_{\mathrm{e}}=n_{\mathrm{e}}/n_{\mathrm{p}}\simeq1+n_{\mathrm{He^{+}}}/n_{\mathrm{H^{+}}}=1+y_{+}$
and $\mu_{+}\simeq1+4\,n_{\mathrm{He^{+}}}/n_{\mathrm{H^{+}}}=
1+4y_{+}$ assuming $n_{\mathrm{He^{++}}}=0$ and denoting $y_{+}=n_{\mathrm{He^{+}}}/n_{\mathrm{H^{+}}}$.

The number of hydrogen ionizing photons per unit time, $Q(\mathrm{H}^{0})$, emitted by a nebula in equilibrium is
given by (Osterbrock and Ferland, \cite{oster06})
       \begin{equation}
             Q(\mathrm{H}^{0})=\epsilon n_{\mathrm{e}}n_{\mathrm{p}}\alpha_{B}V,
       \end{equation}
where $\alpha_{B}$ is the recombination coefficient given by the following equation, taken from Draine
(\cite{draine11}),
       \begin{equation}
             \alpha_{B}=2.54\times10^{-13}T_{4}^{(-0.8163-0.0208\mathrm{ln}T_{4})}\ \mathrm{cm^{3}s^{-1}} \; .
       \end{equation}
The combination of the two previous equations gives us
the radius, $R_{\mathrm{S}}$, which is the radius of the Str\"omgren sphere
       \begin{equation} \label{stromgren radius}
             R_{\mathrm{S}}=3.17\left(\frac{x_{\mathrm{e}}}{\epsilon}\right)^{1/3}
                        \left(\frac{n_{\mathrm{e}}}{100}\right)^{-2/3}T_{4}^{(0.272+0.007\mathrm{ln}T_{4})}
                        \left(\frac{Q(\mathrm{H}^{0})}{10^{49}}\right)^{1/3}            
       \end{equation}
in units of pc. By combining this equation with the equation (\ref{flux H alpha}), with $R=R_{\mathrm{S}}$
(ionization bounded nebula), the rate of emission of hydrogen-ionizing photons for a given $\mathrm{H}\alpha$
flux, in photons per second is
       \begin{equation}
             Q_{0(\mathrm{H\alpha})}=8.59\times10^{55}T_4^{(0.126+0.01\mathrm{ln}T_4)}D^2F_0(\mathrm{H}_{\alpha}) \; .
       \end{equation}

\subsection{Continuum radio emission}

In the radio frequency region, where $h\nu\ll kT$, the Plank law can be written as
      \begin{equation}
           B_{\nu}=\frac{2\nu^{2}kT}{c^{2}},
      \end{equation}
where $B_{\nu}=j_{\nu}/\kappa_{\nu}$, with $j_{\nu}$ and $\kappa_{\nu}$  the emission and absorption
coefficients at a given frequency $\nu$, respectively.

The radio flux density at a distance $\textit{D}$ from the nebula is given by
      \begin{equation}
           S_{\nu}=\frac{L_{\nu}}{4\pi D^{2}},
      \end{equation}
with $ L_{\nu}=\int_V 4 \pi j_{\nu}\epsilon \mathrm{d}V$. 
Assuming that the nebula is an optically thin sphere of radius $\textit{R}$, the radio flux density 
can then be written
      \begin{equation} \label{radio flux}
           S_{\nu}=\left(\frac{4\pi R^{3}}{3D^{2}}\right)\epsilon B_{\nu}\kappa_{\nu} \; .
      \end{equation}

The continuum free-free effective absorption coefficient at radio frequencies is given by (Osterbrock and Ferland, \cite{oster06})
       \begin{equation}
              \kappa_{\nu}=8.24\times10^{-2}T_\mathrm{e}^{-1.35}\nu^{-2.1}n_{+}n_{\mathrm{e}},
       \end{equation}
where the temperature $T_\mathrm{e}$ is in K and the frequency $\nu$ is in GHz. The ion density is equal to 
$n_{+}=n_{\mathrm{H}^{+}}+n_{\mathrm{He}^{+}}=
n_{\mathrm{e}}$.

The ionized mass of the nebula is defined by equation (\ref{ionized mass}). By adopting the same
formalism as in the first part of this Appendix, the ionized mass in solar masses is finally found to be
       \begin{equation} \label{ionized mass radio}
              M_{i(radio)}=5.82\times10^{-5}\frac{1+4y_{+}}{1+y_{+}}T_4^{0.175}
                    \left(\frac{\nu}{4.9}\right)^{0.05}\epsilon^{1/2}\theta^{3/2}D^{5/2}S^{1/2}_{\nu},
       \end{equation}
where $\nu$ is the radio frequency in GHz, $\theta$ is the angular radius of the ionized nebula in arcsec,
$\textit{D}$ is the distance to the nebula in kpc and $S_{\nu}$ is the radio flux density in mJy.

The rate of emission of hydrogen-ionizing photons, for a given radio flux density, can be found by combining equations
(\ref{stromgren radius}) and (\ref{radio flux}):
       \begin{equation}
        Q_{0(\mathrm{radio})}=8.72\times10^{43}T_4^{(-0.466-0.0208\mathrm{ln}T_4)}
               \left(\frac{\nu}{4.9}\right)^{0.1}x_e^{-1}D^2S_{\nu} \; .
       \end{equation}

\section{Photodissociation region}

According to Tielens (\cite{tie05}), the total mass of hydrogen in the PDR is given by 
       \begin{equation}
          M_{\mathrm{H}}=\frac{4\pi D^2 m_{\mathrm{H}}}
                         {X_{\mathrm{C}}A_{\mathrm{ul}}E_{\mathrm{lu}}}
                         \left(\frac{g_{l}}{g_{u}}\mathrm{exp}
                         [E_{\mathrm{lu}}/kT_{\rm PDR}]+1\right)F_{[\ion{C}{ii}]},
       \end{equation}
where $X_{\mathrm{C}}$ is the C/H abundance in number, $T_{\rm PDR}$ the temperature of the gas, 
$g_{l}$, $g_{lu}$, $A_{\mathrm{ul}}$ and
$E_{\mathrm{lu}}$ the statistical weights, transition probability and energy
difference of the levels involved to this transition, $\textit{D}$ the distance
to the star and $F_{[\ion{C}{ii}]}$ the observed $[\ion{C}{ii}]$ 158\ \mbox{$\mu$m}
flux. From the cooling law, we have that (Tielens \cite{tie05})
       \begin{equation}
          n_{\mathrm{u}}A_{\mathrm{ul}}h\nu_{\mathrm{ul}}=\frac{(g_{\mathrm{u}}
                       /g_{\mathrm{l}})\mathrm{exp}[-h\nu_{ul}/kT_{\rm PDR}]}
                       {1+n_{\mathrm{crit}}/n_\mathrm{H^0}+(g_{\mathrm{u}}
                       /g_{\mathrm{l}})\mathrm{exp}[-h\nu_{ul}/kT_{\rm PDR}]+1},
       \end{equation}
where $n_\mathrm{H^0}$ is the number density of atomic hydrogen and the $n_{\mathrm{crit}}$
the critical density, given by $n_{\mathrm{crit}}=
3.2\times10^3T_2^{-0.1281-0.0087\mathrm{ln}T_2}$ cm$^{-3}$ where $T_2 = T_{\rm PDR}/(10^{2}$ K) (Draine
\cite{draine11}). By replacing the known parameters with their values
and by considering that $T_{\rm PDR}\gg
E_{\mathrm{ul}}/k=92 K$, we finally have
  \begin{equation}
          M_{\mathrm{H}}(\mathrm{M}_{\odot}) = 4.93\times10^9
                     \left(1+\frac{n_\mathrm{crit} / n_\mathrm{H^0}}{3}\right)
                     D^2 
                     \left(\frac{F_{[\ion{C}{ii}]}}{X_{\mathrm{C}}}\right) \ ,
      \end{equation}
where $\textit{D}$ is in kpc and $F_{[\ion{C}{ii}]}$ is in $\mbox{W}\ \mbox{m}
^{-2}$.

\end{appendix}

\end{document}